
\input harvmac.tex

\def\inbar{\,\vrule height1.5ex width.4pt depth0pt}
\def\IB{\relax{\rm I\kern-.18em B}}
\def\IC{\relax\hbox{$\inbar\kern-.3em{\rm C}$}}
\def\ID{\relax{\rm I\kern-.18em D}}
\def\IE{\relax{\rm I\kern-.18em E}}
\def\IF{\relax{\rm I\kern-.18em F}}
\def\IG{\relax\hbox{$\inbar\kern-.3em{\rm G}$}}
\def\IH{\relax{\rm I\kern-.18em H}}
\def\II{\relax{\rm I\kern-.18em I}}
\def\IK{\relax{\rm I\kern-.18em K}}
\def\IL{\relax{\rm I\kern-.18em L}}
\def\IM{\relax{\rm I\kern-.18em M}}
\def\IN{\relax{\rm I\kern-.18em N}}
\def\IO{\relax\hbox{$\inbar\kern-.3em{\rm O}$}}
\def\IP{\relax{\rm I\kern-.18em P}}
\def\IQ{\relax\hbox{$\inbar\kern-.3em{\rm Q}$}}
\def\IR{\relax{\rm I\kern-.18em R}}
\font\cmss=cmss10 \font\cmsss=cmss10 at 7pt
\def\IZ{\relax\ifmmode\mathchoice
{\hbox{\cmss Z\kern-.4em Z}}{\hbox{\cmss Z\kern-.4em Z}}
{\lower.9pt\hbox{\cmsss Z\kern-.4em Z}}
{\lower1.2pt\hbox{\cmsss Z\kern-.4em Z}}\else{\cmss Z\kern-.4em Z}\fi}
\def\IGa{\relax\hbox{${\rm I}\kern-.18em\Gamma$}}
\def\IPi{\relax\hbox{${\rm I}\kern-.18em\Pi$}}
\def\ITh{\relax\hbox{$\inbar\kern-.3em\Theta$}}
\def\IOm{\relax\hbox{$\inbar\kern-3.00pt\Omega$}}

\def\CM {{\cal M}}
\def\CR {{\cal R}}

\def\CO {{\cal O}}
\def\p {\partial}

\def\inbar{\,\vrule height1.5ex width.4pt depth0pt}
\def\IB{\relax{\rm I\kern-.18em B}}
\def\IC{\relax\hbox{$\inbar\kern-.3em{\rm C}$}}
\def\IP{\relax{\rm I\kern-.18em P}}
\def\IR{\relax{\rm I\kern-.18em R}}
\def\pb{\bar{\partial}}

\Title{\vbox{\baselineskip12pt\hbox{YCTP-P35-91}}}
{\vbox{\centerline{Exact S-Matrix}
\centerline{for}
\centerline{2D String Theory}}}

\centerline{Gregory Moore, M. Ronen Plesser, and Sanjaye Ramgoolam}
\centerline{Department of Physics}
\centerline{Yale University}
\centerline{New Haven, CT\  06511-8167}
\bigskip
\noindent
We formulate simple graphical rules which allow explicit calculation of
nonperturbative $c=1$  $S$-matrices. This allows us to investigate the
constraint of nonperturbative unitarity, which indeed rules out some
theories. Nevertheless, we show that there is an infinite parameter family
of nonperturbatively unitary  $c=1$ $S$-matrices. We investigate the
dependence of the $S$-matrix on one of these nonperturbative parameters. In
particular, we study the analytic structure, background dependence, and
high-energy behavior of some nonperturbative $c=1$ $S$-matrices. The
scattering amplitudes display interesting resonant behavior both at high
energies and in the complex energy plane.

\Date{Nov. 19, 1991}
\noblackbox

\newsec{Introduction}

The definition of the double scaling limit was a significant
breakthrough in physics because it provided a framework in
which nonperturbative aspects of string theory could be
discussed in a rigorous and precise way
\nref
\brkz{E.Br\'ezin and V. Kazakov,
``Exactly Solvable Field Theories of Closed Strings,''
Phys. Lett. {\bf 236B} (1990) 144.}%
\nref\dgsh{M.R. Douglas and S. Shenker,
``Strings in Less Than One Dimension,''
Nucl. Phys. {\bf B335} (1990) 635.}%
\nref\grmgdl{D. Gross and A. Migdal,
``Non-perturbative Two Dimensional Quantum Gravity,''
Phys. Rev. Lett. {\bf 64} (1990) 127.}%
\refs{\brkz-\grmgdl}. Although the nonperturbative formulation given by the
matrix models is not necessarily the ``correct'' one, it is the only one
available and hence deserves close scrutiny.

Unfortunately, the original hope
that a Painlev\'e I transcendent would provide the nonperturbative
definition of the free energy of pure 2D quantum
gravity evaporated with the discovery that the reality of the free energy
inevitably clashes with ``physical'' constraints
arising from matrix-model/topological field theory Ward identities
\ref\david{F. David,
``Loop equations and non-perturbative effects in 2D Quantum Gravity''
Mod. Phys. Lett. {\bf A5} (1990) 1019.}.
The PI transcendent was subsequently demoted to the status of a
generating functional for the string perturbation series.
This fate has been shared by all solutions to the string equations for
gravity coupled to unitary $c<1$ conformal
matter. Many attempts to circumvent
these difficulties have been made. Those which are
physically well-motivated have proven to be mathematically inconsistent
\ref\dgss{M. Douglas, N. Seiberg,and S.Shenker,
``Flow and Instability in Quantum Gravity''
Phys. Lett. {\bf B244} (1990).}.
Others, while mathematically consistent lack any cogent physical rationale.
We are thus left with the unsatisfying situation of having no nonperturbative
unitary $c<1$ theory of gravity.

There is a two-fold origin of the above dilemma.
First, there is no simple spacetime interpretation of the $c<1$ models.
Hence questions of, e.g.,
whether the specific heat should or should not have an imaginary part
are difficult to resolve. Second and more importantly,
there is no physical principle which isolates a
reasonable parametrization of physically acceptable possibilities.

The first difficulty is not essential thanks to the
$c=1$ model of 2D gravity
\ref\bpiz{E. Br\'ezin, C. Itzykson, G. Parisi, and J.-B. Zuber,
``Planar Diagrams,''
Comm. Math. Phys. {\bf 59} (1978) 35.}.%
\foot{For reviews see
\nref\kazrv{V. Kazakov,
``Bosonic strings and string field theories in one dimensional target
space,'' to appear in the proceedings of the Carg\'ese Workshop
on Random Surfaces, Quantum Gravity, and String Theory, 1990.}%
\nref\klebrev{I. Klebanov,
``String theory in two dimensions,'' Princeton preprint PUPT-1271,
lectures delivered at the ICTP Spring School on String Theory and
Quantum Gravity.}%
\nref\kutrev{D. Kutasov, ``Some properties of (Non) Critical Strings,''
Princeton preprint PUPT-1277,
lectures delivered at the ICTP Spring School on String Theory and
Quantum Gravity.}%
\refs{\kazrv-\kutrev}.}
In the $c=1$ model there is a simple
spacetime interpretation
\ref\joei{J. Polchinski, ``Critical Behavior of Random Surfaces in One
Dimension,'' Nucl. Phys. {\bf B346} (1990) 253.}%
: strings move in two target space
dimensions. The nongauge field theoretic degrees of freedom of the
string are described by a single massless boson field - the
``massless tachyon'' which is related to the eigenvalue density
field of collective field theory
\nref\dj{S.R. Das and A. Jevicki, ``String Field Theory and Physical
Interpretation
of D=1 Strings,'' Mod. Phys. Lett. {\bf A5} (1990) 1639.}%
\nref\wadia{A.M. Sengupta and S.R. Wadia,
``Excitations and interactions in $d=1$ string theory,''
Int. Jour. Mod. Phys. {\bf A6} (1991) 1961.}%
\refs{\dj,\wadia}.
The vertex operator calculations of the $c=1$ matrix model are
Euclidean continuations of the scattering amplitudes of the massless
boson in a half-space, i.e., the ``wall'' $S$-matrix amplitudes in
the terminology of Polchinski
\ref\joefluid{J. Polchinski, ``Classical limit of 1+1 Dimensional
String Theory,'' Texas preprint UTTG-06-91.}.

With this clear physical interpretation there is an equally clear
physical constraint on the theory: the $S$-matrix must be unitary.
That this is true perturbatively might be expected to follow automatically
either from the point of view of collective field theory or from
string perturbation theory. However the question is somewhat more
subtle for nonperturbative definitions of the theory.

In the present paper we study nonperturbative unitarity of the
$c=1$ $S$-matrix. We will show that the situation at $c=1$ is
the reverse of that at $c<1$: while nonperturbative unitarity
may be used to rule out some theories there is a plethora of
nonperturbatively unitary $c=1$ theories.
Thus, the second and more important difficulty
alluded to above will be exacerbated. Indeed, our construction
applies to a wide class of matrix model potentials $V(\lambda)$.
These may be divided into two classes by the asymptotic properties of
$V(\lambda)$ as the eigenvalue variable
 $\lambda\to -\infty$. In theories of type I
$V(\lambda)\to +\infty$ rapidly for $\lambda\to
-\infty$ and $V(\lambda)\to -\lambda^2$ for
$\lambda\to +\infty$. Thus the theory is
effectively defined on a semi-infinite line.
The canonical example of such theories is
defined by  $V(\lambda)\propto -\lambda^2$
on the interval $[A,\infty)$
together with an infinite wall at $\lambda=A$. We will sometimes
specialize to the case $A=0$ where formulas simplify. Although
the potential is not analytic at $\lambda=A$ we expect
that for a smoothed out version of the wall the main results will
be unchanged.  Theories of type II are defined by a smooth potential
$V(\lambda)\to -\lambda^2$ on both ends of the real $\lambda$
axis. Thus, in perturbation theory there appear to be two disconnected
worlds.

In outline, the paper is organized as follows.
In section two we give a definition of the $S$-matrix in terms
of large spacetime asymptotics of correlators of the
eigenvalue density operator of the matrix model. We begin with
an integral representation for the correlators derived in
\ref\moore{G. Moore, ``Double Scaled Field Theory at $c=1$,''
Rutgers/Yale preprint, Nucl. Phys. B, to appear.}
and continue the resulting asymptotics to Minkowski space
to obtain the $S$-matrix. The procedure of this section
is an important technical advance over previous calculations.
For example, in \moore,
the two-, three-, and four-tachyon scattering amplitudes
were obtained to all orders of perturbation theory
using the small length
asymptotics of macroscopic loops. However, at the time
\moore\ was written, extension of the results to more
general amplitudes appeared hopeless. Moreover, while the
results of \moore\ are valid to all orders of perturbation
theory the nonperturbative foundations of these formulae
are shaky. Another advantage of the procedure of section
two is that the nonperturbative formulation of the theory
is unambiguous.

In section three we formulate the result of applying the
procedure of section two in terms of some simple
graphical rules. These rules lead to a relatively
simple and compact form for the $n\to m$ amplitudes
(eq.~(3.6) below). Some particularly simple cases, for
example the nonperturbative $1\to n$ amplitudes are
written out explicitly (eq.~(3.11)). The graphical rules also provide
considerable insight into the fundamental nature of
$c=1$ scattering, making clear how particle production is
possible in a theory of free fermions.

In section four we apply the lessons learned from the
graphical formalism of section three to the problem of
nonperturbative unitarity. Both theories of type I and
theory II are unitary to all orders of perturbation
theory. In section (4.1) we prove that theories of type  I are
nonperturbatively unitary. In section (4.2) we remark that
theory II is not unitary, essentially because the theory
does not take into account fermionic soliton sectors.
Our remark will strike many readers as trivial.
Nevertheless, we think it is important and deserves emphasis.
The unitarity proof for theory I suggests various interesting
generalizations of $c=1$ scattering and leads to a characterization
of a large class of acceptable
nonperturbative definitions of the $c=1$ $S$-matrix.

The exact formulae for $c=1$ scattering amplitudes allow us to investigate
in some detail the analytic structure of the $S$-matrix in section
five. We find several interesting singularities of the analytically-continued
$S$-matrix elements and interpret these in terms of metastable
bound states of matrix model fermions. We emphasize the dependence on
the parameter $A$. This parameter does not appear in perturbative
amplitudes but has an important influence on the nonperturbative answers.

In section six we make some remarks on the worldsheet/Liouville
interpretation of our results. Most importantly, we show that
the vertex-operator-motivated
prescription used in \moore\ to extract correlators from
small loop-length asymptotics of macroscopic loop amplitudes
is equivalent (to all orders of
perturbation theory) to the procedure of section two.

In section seven we make a few remarks on the background-dependence
of the $S$-matrix, focusing on the background dependence of the
singularity structure uncovered in section five. The dependence on
tachyonic perturbations of the background will be addressed in a
separate publication. In section eight we study some aspects of
the high energy behavior of our scattering amplitudes. The amplitudes
exhibit many nontrivial features. While this is hardly surprising from
the close relation of the present $S$-matrix to that of potential
scattering, the interpretation of these features might prove instructive
for string theory. In the conclusion we present some remarks on
directions for future research, and we summarize some technical details
in several appendices.

\newsec{Definition of the $S$-matrix}

\subsec{Eigenvalue Density Correlation Functions}

The Euclidean Green's functions of the
eigenvalue density
$\rho=\psi^\dagger\psi(\lambda,x)$, where $x$ is the
``time'' dimension of the $c=1$ matrix model are defined by
\eqn\gfns{\eqalign{
G_{\rm Euclidean}(x_1,\lambda_1,\dots , x_n,\lambda_n)&\equiv
\langle \mu|T\biggl( \hat\psi^\dagger\hat\psi(x_1,\lambda_1)\cdots
\hat\psi^\dagger\hat\psi(x_n,\lambda_n)\biggr)|\mu\rangle_c\cr
G_{\rm Euclidean}(q_1,\lambda_1,\dots q_n,\lambda_n)&\equiv \int \prod_i dx_i
e^{i q_i x_i}
G(x_1,\lambda_1,\dots x_n,\lambda_n)\ .\cr}
}
Since the
fermions are noninteracting these
Green's functions may be written in terms of the
the Euclidean fermion propagator:
\eqn\propeuc{
S^E(x_1,\lambda_1;x_2,\lambda_2)
=e^{-\mu \Delta x}\int_{-\infty}^\infty {dp\over 2\pi}
e^{-ip\Delta x}I(p,\lambda_1,\lambda_2)
}
where $I$ is the resolvent for the upside down oscillator
Hamiltonian $H=\half p^2-{1\over 8}\lambda^2$. In particular, for
$q>0$
\eqn\iasfg{
I(q,\lambda_1,\lambda_2)
=(I(-q,\lambda_1,\lambda_2))^*
=\langle\lambda_1 |{1\over H-\mu-i q}|\lambda_2\rangle
}

Thus we obtain the integral representation for the
eigenvalue correlators \moore:
\eqn\nptev{\eqalign{
G_{\rm Euclidean}(q_i,\lambda_i)&={1 \over n}
\int \prod_{i=1}^n {d q_i\over 2\pi}
e^{-i q_i x_i}\sum_{\sigma\in\Sigma_n}
\prod S^E(x_{\sigma(i)},\lambda_{\sigma(i)};x_{\sigma(i+1)},
\lambda_{\sigma(i+1)})\cr
&= {1 \over n} \delta(\sum q_i)
\int_{-\infty}^\infty dq\sum_\sigma \prod_{k=1}^n
I(Q^\sigma_k,\lambda_{\sigma(k)},\lambda_{\sigma(k+1)})\cr}
}
where $Q_k^\sigma\equiv q+q_{\sigma(1)}+\cdots q_{\sigma(k)}$.

\subsec{Relation to Collective Field Theory}

We now relate the exact eigenvalue correlators \nptev\ to the
correlation functions of
the Das-Jevicki collective field formulation, as explained in
\nref\lpsflds{G. Moore and N. Seiberg, ``From Loops to Fields in
2D Gravity,'' Rutgers/Yale preprint,
to appear in the Int. Jour. of Mod. Phys.}%
\nref\prev{D. Gross and I. Klebanov,
``Fermionic String Field Theory of $c=1$ 2D Quantum Gravity,''
Nucl. Phys. {\bf B352} (1991) 671.}%
\refs{\wadia,\joefluid,\lpsflds,\prev}.%
\foot{Of course, the connection had also been
extensively discussed in \refs{\wadia,\prev,\joefluid} before \lpsflds.
Nevertheless,
the method of appendix A in \lpsflds\ leads to the
simple calculation of this paper, which we believe is
new.}

First, let us find the large $\lambda$ behavior of $I$.
We work on half of the $\lambda$ axis, change variables
to $\lambda=2\sqrt{\mu}\cosh\tau$ and identify $\delta \rho\equiv
\rho-\langle \rho\rangle\equiv
\p_{\lambda}\chi$ as usual.
If we hold $\tau$ fixed then
asymptotically at large $\mu$ the integral
$I$ has the form of a direct and reflected propagator \lpsflds:
\eqn\expdi{\eqalign{
I(q,\tau_1,\tau_2)
{\buildrel \mu\to\infty\over\sim}
{i\over\sqrt{4\mu\sinh\tau_1\sinh\tau_2}}
\Biggl[e^{-q|\tau_1-\tau_2|}& e^{i\mu| G(\tau_1)-G(\tau_2)|}
D(q,\tau_1,\tau_2)\cr
+i &e^{-q(\tau_1+\tau_2)}e^{i\mu( G(\tau_1)+G(\tau_2))}
R(q,\tau_1,\tau_2) \Biggr]\ .\cr}
}
Here $G(\tau)=\tau-\half\sinh 2\tau$ is the WKB phase factor of a fermion
wavefunction, and $R=1+{i\over\mu} C^{(1)}+{1\over\mu^2}C^{(2)}+\cdots$.
The $C^{(i)}$ are real, polynomials in the $q_i$, and
rational functions of $w_i=e^{\tau_i}$. Finally,
$D(q,\tau_1,\tau_2)=R(q,\tau_1,-\tau_2)$
for $\tau_1>\tau_2$.

The correlation functions of the Das-Jevicki collective
field theory are obtained by
inserting \expdi\ into \nptev, expanding the product, and keeping
only
those terms for which the nonperturbatively oscillating factors
$e^{\pm i \mu G(\tau)}$ cancel. This defines a new set of
Green's functions $\tilde G(q_i,\tau_i;\mu)$ (at least as
asymptotic expansions in $1/\mu$) which are the Euclidean
continuation of the
Green's functions of the Das-Jevicki theory to all orders
of perturbation theory:
\eqn\djconn{
\tilde G_{\rm Euclidean}(\tau_i,q_i;\mu)
=\prod_i {1\over 2\sqrt{\mu}\sinh\tau_i}
\langle 0|\prod_i \p_\tau \chi_i|0\rangle_c\ .
}

Although in principle this procedure gives a straightforward
method for calculating the off-shell Green's functions of
$\p_\tau\chi$, in practice it becomes
cumbersome already at one-loop
\ref\mrsh{The one-loop off-shell propagator was calculated in
this way by G. Moore and S. Shenker; unpublished.
The procedure described in the text is {\it ad hoc}, but is necessary to
reproduce even the perturbative expansion. It is related to
the neglect of similarly nonperturbatively oscillating terms
in \refs{\wadia,\prev}. While a better understanding is certainly called for,
we will not address the problem in more
detail here since in any case the nonperturbatively oscillating terms
do not contribute to the $S$-matrix.}.
Nevertheless, as we will see,
it is an extremely useful observation for extracting the
$S$-matrix.

\subsec{Large $\lambda$ Asymptotics}

Our strategy for extracting $S$-matrix elements (in Minkowski space) and
(microscopic) vertex operator correlators (in Euclidean space)
will be to extract
the coefficients of appropriate terms in the large $\lambda_i$
asymptotic expression of $\tilde G$.
{}From the collective-field-theory we may invoke a coordinate space
version of the LSZ reduction procedure to
identify the
$S$-matrix as the coefficient of the term multiplying
appropriate incoming and outgoing wavefunctions of $\tau$
\nref\fadd{L. Faddeev, in {\it Methods in Field Theory} Proc.
of the Les Houches Summer School 1975, Balian et. al. eds.
World Scientific}%
\nref\itzub{C. Itzykson and J.B. Zuber, {\it Quantum Field Theory},
Mc Graw-Hill, 1980.}%
\refs{\fadd,\itzub}.

In the $\tau\to\infty$ limit the complicated rational
expressions in \expdi\ simplify drastically.
Since \iasfg\ is just the Green's function for $H$ it may itself
be expressed in terms
of parabolic cylinder functions. From this
representation one obtains the asymptotics
for $\lambda_i\to \pm \infty$. The calculation is outlined in appendix A.
For $\lambda_i\to+\infty$ we obtain
\eqn\asympi{\eqalign{
I(q,\lambda_1,\lambda_2)\sim {-i\over\sqrt{\lambda_1\lambda_2}}
\biggl[e^{-q|\tau_1-\tau_2|}e^{i\mu|G(\tau_1)-G(\tau_2)|}&
\cr
+R_q e^{i\mu(G(\tau_1)+G(\tau_2))} &e^{-q(\tau_1+\tau_2)}\biggr]
\bigl(1+\CO(e^{-\tau_i})\bigr)\qquad q>0\cr
I(q,\lambda_1,\lambda_2)\sim {i\over\sqrt{\lambda_1\lambda_2}}
\biggl[e^{q|\tau_1-\tau_2|}e^{-i\mu|G(\tau_1)-G(\tau_2)|}&
\cr
+(R_q)^* e^{-i\mu(G(\tau_1)+G(\tau_2))} &e^{q(\tau_1+\tau_2)}\biggr]
\bigl(1+\CO(e^{-\tau_i})\bigr)\qquad q<0\ .\cr}
}

The ``Euclidean fermion reflection matrix'' $R_q$ will play a key
role in what follows.
In theories of type I, defined on $\lambda\in [A,\infty)$ we find
\eqn\gamhf{
R_q=ie^{i\mu log\mu}\mu^{-|q|}\sqrt{1+i e^{-\pi(\mu+i|q|)}\over
1-i e^{-\pi(\mu+i|q|)}}
\sqrt{\Gamma(\half-i\mu+|q|)\over\Gamma(\half+i\mu-|q|)}
\ .}
for $A=0$. For nonzero $A$ the result is given in
equation $(A.7)$.

In theory II we have
\eqn\gmfrm{R_q=-{i\over\sqrt{2\pi}}e^{i\mu log\mu}\mu^{-|q|}
e^{{\pi\over 2}(\mu+i|q|)} \Gamma(\half-i\mu+|q|)\ .}
The factors $R$ in \gamhf\ and \gmfrm\ agree to all orders of
perturbation theory
\eqn\difference{
R_q^{II}={1\over 1+i e^{-\pi(\mu+i|q|)}}R_q^{I}
}
and have a perturbative expansion of the form $R_q\sim 1 + \CO(1/\mu)$.
In theory II there are two worlds and we must also calculate the
``transmission''
probabilities obtained from the asymptotics of $I$ for
$\lambda_1=2\sqrt{\mu}\cosh\tau_1\to+\infty$ and
$\lambda_2=-2\sqrt{\mu}\cosh\tau_2\to-\infty$. In this case the ``direct
propagator'' in \asympi\ does not appear and the transmitted
amplitude is
\eqn\trans{T_q=-i e^{-\pi\mu-i\pi|q|}R_q\ .}

We may now derive the large $\lambda$ asymptotics of the collective
field theory Green's function $\tilde G$.
Substitute \asympi\ into \nptev\ and expand,
keeping only terms for which factors of $e^{i\mu G(\tau)}$ cancel.
For
large $\tau$ the two-point function behaves like:
\eqn\twpt{
\tilde G_{\rm Euclidean}= \delta (q_1 + q_2)
{1\over 4\mu\sinh\tau_1\sinh\tau_2}
\Biggl[|q|e^{-|q||\tau_1-\tau_2|}
+ \CR_2(q_1,-q_1;\mu)e^{-|q|(\tau_1+\tau_2)}\Biggr]
}
giving a direct and reflected propagator for the string-theoretic
massless tachyon.
Specifically, for $q > 0$
\eqn\rtwo{
\CR_2(q,-q;\mu) = \int_0^q dx R_x R_{q-x}^*
}
For three or more operators the resulting expression for $\tilde G$
will have large $\tau_i$ asymptotics:
\eqn\defrn{
 \tilde G\sim\delta(\sum q_i)\prod_{i=1}^n{e^{-|q_i|\tau_i}\over
2\sqrt{\mu}\sinh\tau_i}\CR_n(q_1,\dots,q_n;\mu)
\biggl[1 + \CO(e^{-\tau_i})\biggr]
\ .}
Equation \defrn\ may be taken as the definition of the
functions $\CR_n$.
In theory II we define in an analogous way
the functions $\CR_n(\epsilon_i,q_i)$
where $\epsilon_i=\pm$ for $\lambda_i\to \pm \infty$.

\subsec{Continuation to Minkowski Space}

The analytic continuation to Minkowski space is most clearly
understood by starting with the Minkowski space formulation
of the Das-Jevicki theory, continuing that theory to Euclidean
space and comparing with the Euclidean free fermion amplitudes.
By studying the behavior of propagators we can obtain the
following analytic continuation prescription. Consider
the function $\CR_n(q_1,\dots, q_k;q_1'\dots q_l')$
in some kinematic region where
$q_i>0$ and $q_i'<0$. We replace $q_i\to -i \omega_i$
and $q_i'\to i\omega'_i$, or, briefly, $|q|\to -i\omega$.
Here and hereafter $\omega$ represents a real positive
number, the energy of a massless tachyon.

Explicitly, in theory I with $A=0$ the reflection factor
becomes a pure (energy-dependent) phase:
\eqn\rgmnk{\eqalign{
R_q\to
&e^{i\mu log\mu}\mu^{i\omega}i\sqrt{1+i e^{-\pi(\mu+\omega)}\over
1-i e^{-\pi(\mu+\omega)}}
\sqrt{\Gamma(\half-i(\mu+\omega))\over\Gamma(\half+i(\mu+\omega))}\equiv
ie^{i\mu log\mu}\mu^{i\omega}e^{i\Theta(\mu+\omega)}\qquad q>0\cr
R_q\to &i e^{i\mu log\mu}\mu^{i\omega}
e^{-i\Theta(\mu-\omega)}\qquad\qquad\qquad q<0\ .\cr}
}
More generally the analytic continuation will define a phase
$e^{i\Theta(\mu+\omega,A)}$.
The factors of $i e^{i\mu log\mu}$ and $\mu^{i\omega}$ do not affect final
amplitudes. The former cancels out of eq.~(3.7) below while the
latter may be absorbed as a phase redefinition of the states.

In theory II $R$ does not become a pure phase but has absolute
magnitude $(1+e^{-2\pi(\mu+\omega)})^{-1/2}$. Similarly $T$
has magnitude $(1+e^{2\pi(
\mu+\omega)})^{-1/2}$. For $\mu$
large and positive any
$S$-matrix element involving $k$ bosons traversing the barrier
is exponentially suppressed and of order $e^{-2\pi k \mu}$.

\subsec{Definition of the $S$-Matrix}

In theory I
there is an incoming and outgoing Fock space of bosonic states
defined by oscillators $\alpha(\omega)$ and normalized by
\eqn\norml{[\alpha(\omega),\alpha^\dagger(\omega')]=\omega
\delta(\omega-\omega')\ .}
In the space of incoming states $\CH^{in}$, $|\omega_1,...\omega_k\rangle$
represents
a set of left-moving bosons approaching the wall from the left with
wavefunction
\eqn\incmr{
\psi^L_{\omega_1,\dots \omega_k}(t_i,\tau_i)={1\over \sqrt{k!}}
\sum_\sigma \prod e^{-i\omega_{\sigma(i)}(t_i+\tau_i)}
\ .}
In the outgoing space
$\langle \omega|$ represents a set of outgoing right-moving bosons with
wavefunction
\eqn\outgl{
\psi^R_{\omega'_1,\dots \omega'_l}(t'_i,\tau'_i)=
{1\over \sqrt{l!}}\sum_\sigma \prod e^{-i\omega_{\sigma(i)}'(t'_i-\tau'_i)}
\ .}
The wavefunctions $\psi^R,\psi^L$ are continuations of the
Euclidean wavefunctions $\prod_i e^{-|q_i|\tau_i + i q_i x_i}$. Thus,
the analytic continuation of the previous subsection shows
that negative and positive $q$ correspond to incoming and
outgoing particles respectively.

Putting together \defrn\ with \incmr\outgl\ we may write
the large $\tau$ asymptotics of the Minkowski space
Green's function as:
\eqn\minkgf{\tilde G_{\rm Minkowski}(\omega,\tau)\sim
\delta(\sum_{i=1}^k\omega_i-\sum_{i=1}^l\omega'_i)
\prod_{i=1}^k\bigl(\psi^L_{\omega_i}(t_i,\tau_i)\bigr)^*
\prod_{i=1}^l\bigl(\psi^R_{\omega'_i}(t_i',\tau_i')\bigr)
\sqrt{k! l!} S_c(\omega_i|\omega'_i)
}
where $S_c$ is obtained from
${i^{k+l} \over \sqrt{k! l!}} \CR_n$ by analytic continuation.

Equation \minkgf\ defines the connected $S$-matrix element:
\eqn\defsc{
\langle \omega_i|S|\omega_i'\rangle=\delta (\sum\omega_i-\sum\omega_i')
S_c(\omega_i|\omega'_i) +
{\rm disconnected\ terms}
\ .}

\newsec{Calculation of the $S$-matrix}

\subsec{Diagrammatic Formalism}

The procedure of extracting terms for which $e^{\pm i \mu G(\tau)}$
cancel from the product of the functions
$I$ can be given a diagrammatic interpretation which facilitates
calculation and which will be crucial to our proof that theory I
is unitary.%
\foot{The formalism is closely related to
old-fashioned non-covariant perturbation
theory for the fermions. Several special
features of our problem lead to {\it ad hoc}
rules which necessitate our detailed discussion below.}

As in a Feynman diagram there is a vertex in the $(x,\tau)$ half-space
corresponding to each operator $\psi^\dagger\psi(x,\tau)$.
While the final result will of course be independent of the order in which
the $\tau_i$ are increased to infinity, in intermediate steps we will
choose some order and locate the vertices accordingly.
Points are connected by line segments, representing the integral
$I$, to form a one-loop graph. Since the expression for $I$ in
\asympi\  has two terms and we have both direct
and reflected propagators as in
\fig\idiagram{a.) A pictorial version of the integral $I$ for
positive momentum. (b) A pictorial version of the integral $I$ for
negative momentum}.
Each line segment carries a momentum and an arrow. Note that
in \idiagram\ the reflected propagator, which we call simply
a ``bounce,'' is composed of two segments with opposite
arrows and momenta. These line segments are joined to form
a one-loop graph according to the following rules:

\medskip

{\parindent=1truein
\item{RH1.} Lines with positive (negative) momenta slope upwards to the
right (left).

\item{RH2.} At any vertex arrows are conserved and momentum
 is conserved as
time flows upwards. In particular momentum
$q_i$ is inserted at the vertex
as in
\fig\vertex{Incoming and outgoing vertices.
The dotted line carrying negative (positive) momentum $q_i$ should be
thought of as an incoming (outgoing) boson with energy $|q_i|$.
Momentum carried by lines is always conserved as time flows upwards.}.

\item{RH3.} Outgoing vertices at $(x_{out},\tau_{out})$ all have
later times than incoming vertices $(x_{in},\tau_{in})$: $x_{out}>x_{in}$.

}

\medskip

Since diagrams drawn according to these rules correspond to
real processes taking place in spacetime they will be called real
histories. Some examples are shown in
\fig\onetwo{The possible real histories for $1\to 2$ scattering}
and
\fig\twotwo{Some possible real histories for $2\to n$ scattering}.
There is a finite number (less than $n!$) of ways of connecting
the dots. Since $\sum q_i=0$ there is an overall undetermined
loop momentum $q$, however the constraints RH1-RH3 restrict $q$
to lie in a finite interval.

To each real history we associate an amplitude, easily derived
from \asympi\ and illustrated in
\fig\eclrl{Diagrammatic rules in Euclidean space}
and
\fig\mnkrl{Diagrammatic rules in Minkowski space}.
The total amplitude is simply obtained by summing over
real histories to produce a formula which reads, schematically,
\eqn\radmis{\CR=i^n\sum_{RH}\pm \int dq \prod_{\rm bounces} R_Q (-R_Q)^*}

In theory II we take two copies of $(x,\tau)$
space. Transmission amplitudes are computed by connecting vertices
in one half space to another. Each transmission line is weighted
with a factor of $T_q$. Note in particular that since two fermions
make a boson every transmission of a boson has a nonperturbative
suppression of $e^{-2\pi \mu}$.

\subsec{Explicit Formula}

We now write the equation \radmis\ more precisely. Each real history
has the structure of a series of ``brackets''
\fig\bracket{Typical bracket configurations}
separated by bounces since (by considering momentum
conservation) direct propagators can only connect
incoming to incoming and outgoing to outgoing vertices.
Each bracket must contain at least
one insertion of momentum, so in an $n\to m$ amplitude there
can be up to $Min\{n,m\}$ incoming and outgoing brackets.
The number of incoming brackets equals the number of outgoing
brackets. As we traverse the loop,
we encounter an increasing set of
momentum insertions. The factors for each bounce only
depend on the net momenta flowing through the bracket, but
there is a combinatoric factor counting the number
of ways of forming a bracket out of a given set $T$ of
momenta. Consider for example the case where $T$ is a set of
positive momenta. The only constraints on the formation of
a bracket as in \bracket\ is that, if $q^*$ corresponds to
the vertex $\tau_*$ with the largest $\tau$ in $T$ then
the direct propagators immediately before and after the
insertion of $q^*$ must have positive and negative
momenta, respectively. Let $T_1$ be the momenta inserted
before the insertion of $q^*$ and $T_2$ the momenta inserted
after. The sum over real histories involves a
sum with weights:
\eqn\brkti{\eqalign{
\sum_{T_1\amalg T_2\amalg\{q^*\}=T}\theta(q(T_2)+q^*-Q)
\theta(Q-q(T_2))(-1)^{|T_2|}
&\qquad\qquad\qquad\qquad\cr
=\sum_{T_1\amalg T_2\amalg\{q^*\}=T}\bigl(\theta(Q-q(T_2))
&-\theta(Q-q(T_2)-q^*)\bigr)(-1)^{|T_2|}\cr
&=\sum_{T_1\amalg T_2=T}(-1)^{|T_2|}\theta(Q-q(T_2))\cr
&=-\sum_{T_1\amalg T_2=T}(-1)^{|T_2|}\theta(q(T_2)-Q)\cr
&\equiv f_+(T,Q)\cr}
}
where the notation $q(S)$ means the sum of momenta in a set
$S$ and in going from the second to the third expression
we have noticed that the two terms in
the sum correspond to subsets which do and do not include $q^*$.
Note in particular that the final weighting factor makes no reference
to $q^*$ and is therefore independent of $\tau$-ordering.
Similarly, for a set $T$ of negative momenta as in \bracket\ we have
a weighting factor
\eqn\brktii{\eqalign{
f_-(T,Q)=&\sum_{T_1\amalg T_2=T}\theta(q(T_1)-Q)(-1)^{|T_2|}\cr
&=-\sum_{T_1\amalg T_2=T}\theta(Q-q(T_1))(-1)^{|T_2|}\ .\cr}
}
Note that the form of $f$ differs for positive and negative momenta.
For convenience we set $f_{\pm}(\emptyset,Q)=0$.

We are now in a position to write equation \radmis\ precisely.
Let $S=\{q_1,\dots,q_n\}$ be the set of momenta,
$S^{\pm}$ the set of
positive and negative momenta, and define an admissible
filtration (AF) of order $k$ to be a tower of subsets of momenta:
\eqn\filt{\emptyset\equiv F_0\subset F_1\subset F_2\subset\cdots\subset
F_{2k}=S
}
such that
\eqn\diffcond{\eqalign{
F_j-F_{j-1}\subset S^+\qquad j&=0 \pmod 2 \cr
F_j-F_{j-1}\subset S^-\qquad j&=1 \pmod 2\ .\cr}
}
Then we have the (Euclidean space) ``filtration formula''
\eqn\fltfrm{
\CR_n=i^n \sum_{k=1}^m
{1\over k} \sum_{AF_k}
\int d Q\prod_{j=1}^{2k} f_{(-1)^j}(F_j-F_{j-1},Q+q(F_j))
\prod_{j=1}^{k} R_{Q+q(F_{2j-1})}
R_{Q+q(F_{2j})}^*
}
where $m = {\rm Min} \{|S^+|,|S^-|\}$.
The combinatoric factors $f_{\pm}$ impose kinematic constraints
which render the $Q$-integral finite.

Examining the reflection factors of \gamhf ,\gmfrm\ we see that the $\mu$
dependence of $R_q$ is only through the combination $\mu + i|q|$,
except for the
(noncancelling) prefactor $\mu^{-|q|}$.
This observation allows us to compute the derivative
$\p_{\mu} \big(\mu^{\half\sum|q_i|}\CR\big)$ by converting it to a $Q$
derivative. This has the effect of killing
the loop integration, extracting a boundary term wherever
the argument of one of the theta functions vanishes. Thus we can
write a formula
\eqn\radmisi{
{\p\over\p\mu}\biggl(\mu^{\half\sum|q_i|}\CR_n\biggr)=
i^{n+1} \mu^{\half\sum|q_i|}
\sum_{RH'}\pm \prod_{\rm bounces} R_Q R_Q^*
}
where the real histories allowed in \radmisi\ have the added condition

{\parindent=1truein

\item{RH4.} Exactly one direct or reflected propagator carries
zero-momentum. Direct propagators with zero momentum have a factor $+1$.

}

The following useful properties
are immediate consequences of these expressions.
The functions $\CR_n(q_1,\dots,q_n;\mu)$ are real and
totally symmetric functions of the $q_i$ and are
defined on the
plane $\sum q_i=0$ in $\IR^n$. If $S$ is any subset of
momenta then the functions are continuous but have discontinuous
derivatives across the
hyperplanes $H_S$ defined by $q(S)\equiv \sum_{q\in S} q=0$.
On the complements of these hyperplanes the $\CR_n$ are analytic.
Indeed, we can Taylor-expand for small $q_i$ and
the Taylor-coefficients are polynomials in
polygamma functions (see e.g. appendix C), although the
coefficients in the Taylor expansion change across $H_S$.
Similarly, in the asymptotic expansion of $\CR_n$ for
$\mu\to +\infty$ the coefficients
of $1/\mu^n$ are polynomials in the $q_i$, although the
polynomial changes across the hypersurfaces $H_S$.
The analytic structure of $\CR_n$ will be discussed
further in section 5 below.

\subsec{A Low-Energy Theorem}

Suppose any momentum $p_*$ is taken to zero. Physically the vertex operator
inserting $p_*$ becomes proportional to the cosmological constant and the
resulting limit should be expressed as a derivative with respect to $\mu$ of
an $n-1$ point function. This can be proven by noting that if
$p_*\in F_j- F_{j-1}$ then
\eqn\limfctor{f_\pm(F_j-F_{j-1},Q)\to
-p_* {\p\over\p Q} f_\pm(F'_j-F_{j-1},Q)}
where $F'_j=F_j-\{ p_*\}$. Hence combining this remark with \radmisi
\eqn\lmamp{\mu^{\half\sum|q_i|}\CR_n\to
p_*{\p\over\p \mu}\biggl(\mu^{\half\sum
|q_i|}\CR_{n-1}\biggr)
}
as any momentum $p_*$ goes to zero.

\subsec{Special Cases}

The general formula \fltfrm\ is somewhat awkward to work with. We give the
explicit formula in some special cases.

1. If $S^+=\{q_1,\dots q_{n-1}\}$, $q_n<0$. Then we have:
\eqn\oton{
{\p\over\p\mu}
\biggl(\mu^{|q_n|}\CR\biggr)=i^{n+1}\sum_{S_1\amalg S_2=S^+}(-1)^{|S_2|}
R_{q(S_1)}R_{q(S_2)}^*
\ .}

Correspondingly, in theory I the $S$-matrix element for an
incoming state $|\omega\rangle$ to produce
$\langle \omega_1,\dots,\omega_{n-1}|$
is given by
\eqn\prtprd{
S(\omega|\omega_i)={i^{n+1}\over \sqrt{(n-1)!}}
\sum_{S_1\amalg S_2=S^+}(-1)^{|S_2|}
\int_0^{\omega(S_2)} d x
e^{i\Theta(\mu+\omega-x)
+i\Theta(x-\mu)}
\ .}
This allows us to compute the probabilities for particle production when an
incoming tachyon impinges upon the wall. A discussion of this effect follows in
section 8.

We also note that the content of \oton\ and \prtprd\ can be expressed
in terms of {\it linear} Ward identities on the amplitudes,
which could be interesting in Liouville theory.

2. We can use our rules to write a relatively
simple result for the $S$-matrix for scattering
$\omega_1+\omega_2\to \sum \omega_i'$. This consists of two terms
with either 2 or 4 bounces.
Let $S^+ = \{ q_1,\ldots q_{n-2}\}$, and $S^- = \{ q_{n-1},q_n\}$, then:
\eqn\twon{\eqalign{
\CR_n=\sum_{T_1 \amalg T_2 = S^+}
&(-1)^{|T_2|} \int_0^{q(T_2)} \big( 1 - \theta (Q + q_{n-1}) -
\theta (Q+q_n) \big) R_Q^* R_{Q + q(S^-)} dQ \cr
+ \half \sum_{S_1 \amalg S_2 = S^+} &\int_0^{q(S_2)}
f_+(S_2,Q) R_{q(S_2) - Q} R_Q^*
\biggl[ f_+\big(S_1,Q + q_n + q(S_1) \big)
R_{q_n+Q} R_{Q + q_n + q(S_1)}^* \cr
&\qquad \qquad + f_+\big(S_1,Q + q_{n-1} + q(S_1) \big)
R_{q_{n-1}+Q} R_{Q + q_{n-1} + q(S_1)}^* \biggr] \ .\cr
}}

Specializing further,
consider $2\to 2$ scattering. Let $n=4$ above, and take
$q_1+q_3>0$, $q_1+q_4>0$. We find:
\eqn\twotwo{\eqalign{
{\p\over\p\mu}\biggl(\mu^{\half\sum|q_i|}\CR_4\biggr)=
&-4\mu^{\half\sum|q_i|}
Im\biggl[R_{q_1+q_2}R_0^*-R_{q_1}R_{q_2}^*\cr
&-R_{q_2}R_{q_2+q_3}^*R_{q_4}R_0^*
-R_{q_2}R_{q_2+q_4}^*R_{q_3}R_0^*\biggr]\ .\cr}
}
The previously published result for $2\to 2$
scattering in \moore\
involved
an infinite sum of gamma functions
and was rather unwieldy. This illustrates nicely that the formulae
of this paper are a substantial improvement upon those given in
\moore. (And also that there are some remarkable identities on
gamma functions, this particular case is written out in appendix
B.)

\newsec{Unitarity of the $S$-matrix}

\subsec{Theories of Type I are Unitary}

The essential idea of the unitarity proof is to regard the
time evolution of the real histories of the previous section
as a composition of three maps: fermionization, free fermion
scattering, and bosonization: $i_{f\to b}\circ S_{ff}\circ i_{b\to f}$
as in
\fig\composition{A real history as a composition of three maps}.

The most complicated map is the bosonization map, although this
is well-known. We will let $a^\dagger(\nu),a(\nu)$ be delta function
normalized fermion creation operators acting on the Fermi sea:
\eqn\defsea{\eqalign{
a(\xi,+)|\mu\rangle\equiv a^\dagger(\mu+\xi)|\mu\rangle &= 0 \qquad \xi>0\cr
a(\xi,-)|\mu\rangle\equiv a\phantom{{}^\dagger}
(\mu-\xi)|\mu\rangle &= 0 \qquad \xi>0\ .\cr}
}
An operator in the fermionic theory is normal ordered if
it is a sum of terms of the form
$\prod_i a(\xi_i,\epsilon_i)$ with $\epsilon_i \xi_i<0$.
Under bosonization a one-particle incoming state is mapped according to
\eqn\bosnztion{\eqalign{
i_{b\to f}:|\omega\rangle &\to
\int_{-\infty}^\infty d\xi a(\mu+\xi)
a^\dagger(\mu+\xi-\omega)|\mu\rangle\cr
&=\int_0^\omega d\xi a(\mu+\xi)a^\dagger(\mu+\xi-\omega)|\mu\rangle\ .\cr}
}
Note that if incoming states are normalized according to
$\langle \omega|\omega'\rangle=\omega \delta(\omega-\omega')$ then
the map $i_{b\to f}$ is an isometry. This isometry may be
extended to the entire Fock space by
\eqn\bsnztni{
i_{b\to f}:|\omega_1\rangle\otimes\cdots\otimes|\omega_n\rangle\to
\int_{-\infty}^\infty \prod_id\xi_i
\prod_i^n a(\mu+\xi_i)a^\dagger(\mu+\xi_i-\omega_i)|\mu\rangle
\ .}
Note that the operator acting on $|\mu \rangle$
on the RHS is not normal-ordered and hence
the state does not have definite particle-hole number.
We may now imagine normal-ordering the
operators in \bsnztni . To each normal-ordered fermionic monomial we
can associate a diagram of the type drawn in our real histories.
In particular, the
state $\prod_i a(\xi_i,\epsilon_i)|\mu\rangle$ created by a normal-ordered
operator
is associated with a set of incoming bounce-lines carrying negative
momentum $\epsilon_i\xi_i$ and having upward(downward) - pointing arrows
for $\epsilon=\pm$, respectively. By an inductive argument one
can show that the normal-ordering process corresponds exactly to
forming all possible incoming bracket structures discussed in
subsection 3.2 above. The details of this argument appear in appendix D.

Thus the first third of the set of real histories
corresponds in a precise sense to the fermionization of the
incoming bosons. The second stage corresponds to free fermion
scattering. In theory I this scattering is diagonal in the
momentum basis $\prod_i a(\xi_i,\epsilon_i)|\mu\rangle$ and
merely reverses the signs of all $\epsilon_i$. In particular,
\eqn\sff{
\langle \mu | \prod_{i=1}^n a(\xi_i,\epsilon_i) S_{ff}
\prod_{j=1}^m a(\xi'_j,\epsilon'_j) | \mu\rangle =
\delta_{n,m} \sum_{\sigma \in \Sigma_n}
\prod_{j=1}^n \delta (\xi_j + \xi'_{\sigma(j)})
\delta_{\epsilon_j,\epsilon'_{\sigma(j)}}
e^{-i \epsilon_j \Theta(\mu + \xi_j)} \ .
}
The final third of the map is just the inverse of the isometry
$i_{b\to f}$, since our graphical rules implement this inverse
(simply consider reading the diagram from top to bottom). We
have written the bosonic $S$-matrix as a composition of two isometries
and a unitary free-fermion $S$-matrix; it is therefore nonperturbatively
unitary.

\medskip

\noindent{Remarks:}

1. The above discussion allows us to
characterize a large set of nonperturbatively unitary $c=1$
$S$-matrices. Note that at no point were any special properties of the
function $\Theta(x)$ employed: the unitarity equations may be
checked without resort to any identities on the Bernoulli
numbers. The specific form of $\Theta$ only enters in
the comparison with string perturbation theory. Thus we
may construct a nonperturbatively unitary string $S$-matrix
(in two dimensions!) using {\it any}
real-valued function $\Theta(x)$ which has an asymptotic expansion
\eqn\asymreq{\eqalign{
\Theta(x) &\sim arg \Gamma(\half-ix)\cr
&\sim -x log x+x + \sum_{n=1}^\infty {(-1)^n B_{2n}\over 2n(2n-1)}
(1-2^{-(2n-1)})x^{-(2n-1)}\cr}
}
for $x\to +\infty$.
For example, putting the infinite wall at a position $\lambda=A$
leads to the reflection coefficient given in eq. (A.7) below.
For any $A$ this coefficient defines a distinct $S$-matrix,
corresponding to a $\Theta(x,A)$ satisfying \asymreq.
Indeed, by ``inverse scattering'' we expect one can
assign a matrix model potential $V(\lambda;\Theta)$ to any
function $\Theta$ satisfying \asymreq. It would be interesting
to describe the set of potentials which are characterized by
the requirement \asymreq, since, according to the philosophy
of section 7 below, this ambiguity corresponds to the existence of
differences of background geometries detectable only nonperturbatively.

2. The fermionic formulation partially clarifies the question of the $W_\infty$
symmetry which the system is known to respect
\nref\wi{A. Gerasimov, A. Marshakov, A. Mironov, A. Morozov,
and A. Orlov, ``Matrix Models of 2D gravity and Toda Theory,''
Lebedev preprint;
``Matrix Models as Integrable Systems: From Universality to Geometrodynamical
Principle of String Theory,''
Lebedev preprint.}%
\nref\wii{I. Bakas and E. Kiritsis, in {\it Common Trends in
Mathematics and Quantum Field Theories,} Proceedings of the 1990 Yukawa
International Seminar, Prog. Theor. Phys. Supp. {\bf 102};
Berkeley preprint UCB-PTH-90/32.}%
\nref\wiii{M. Fukuma, H. Kawai, and N. Nakayama,
Tokyo preprint UT-562\semi
T. Yoneya, UT-Komaba preprint (1991).}%
\nref\wiv{
M.A. Awada and S.J. Sin, ``Twisted $W_{\infty}$ Symmetry of the KP Hierarchy
and the String Equation of $d=1$ Matrix Models,'' Florida preprint
UFIFT-HEP-90-33; ``The String Difference Equation of $d=1$ Matrix Model and
$W_{1+\infty}$ Symmetry of the KP Hierarchy,''
Florida preprint UFIFT-HEP-91-3.}%
\nref\wv{H. Itoyama and Y. Matsuo,
``$W_{1+\infty}$-Type Constraints in Matrix Models at Finite N,''
Stony Brook preprint ITP-SB-91-10.}%
\nref\wvi{A.M. Semikhatov, ZhETF Lett. {\bf 53} (1991) 12;
``Virasoro Algebra Action on Integrable Hierarchies and Virasoro
Constraints in Matrix Models,'' preprint.}%
\nref\wvii{D. Minic, J. Polchinski, and Z. Yang, ``Translation-Invariant
Backgrounds in 1+1 Dimensional String Theory,''
Texas preprint UTTG-16-91.}%
\nref\wviii{J. Avan and A. Jevicki,
``Classical Integrability and Higher Symmetries of Collective Field
Theory,'' Brown preprint BROWN-HET-801;
``Quantum Integrability and Exact Eigenstates of the Collective String
Field Theory,'' BROWN-HET-824.}%
\nref\wix{S.R. Das, A. Dhar, G. Mandal, S. R. Wadia,
``Gauge Theory Formulation of the c=1 Matrix Model: Symmetries and Discrete
States,'' IAS preprint IASSNS-HEP-91/52.}%
\nref\wx{E. Witten, ``Ground Ring of Two Dimensional String Theory,''
IAS preprint IASSNS-HEP-91/51.}%
\nref\wxi{I.R. Klebanov and A.M. Polyakov, ``Interaction of Discrete States
in Two-Dimensional String Theory,'' Princeton preprint PUPT-1281.}%
\refs{\wi-\wxi,\lpsflds}.
This is not
manifest as a symmetry of the $S$-matrix we have computed.
{}From the point
of view of the free fermion system, the Cartan subalgebra of $W_\infty$
generated by $\CO^{2s,0}$ is a manifest symmetry of \sff, corresponding to the
infinite set of conserved charges $Q_{2s} = \sum p^{2s}$ constructed from the
fermion momenta (note that due to the reflection off the wall, only even powers
are conserved). The realization of the rest of the symmetry algebra is not
obvious. We now see why the symmetry is so deeply hidden in the bosonic
formalism -- the conserved charges do not commute with tachyon number.

3. The above formalism also makes clear how
it is that one can have particle
production in a ``free'' theory. The essential point is
that the fermions scatter with an
energy-dependent phase $e^{i \Theta(E;V)}$.
Thus a particle-hole pair corresponding to
a single boson scatters into a ``dispersed''
particle-hole pair with overlap on states of arbitrary
boson number. This may be regarded as a nonperturbative
generalization of the dispersion of tachyon
wavepackets described in \joefluid.

4. The above discussion suggests an obvious yet interesting
generalization of the present $S$-matrix. We may imagine
adding ``flavor'' quantum numbers to our bosons and fermions
and may further replace the crossings of fermion lines
``in the bulk'' by any factorizable $S$-matrix  compatible
with the bounce factors. The compatibility conditions
have been investigated in
\nref\cherednik{I. Cherednik, ``Notes on Affine Hecke Algebras,''
Bonn-HE-90-04.}%
\nref\gervais{E. Cremmer and J.-L. Gervais, ``The Quantum Strip: Liouville
Theory for Open Strings,'' LPTENS 90/32.}%
\refs{\cherednik,\gervais}.
The above remarks also provide a general
mechanism by which one could use factorizable $S$-matrices
to write nontrivial $S$-matrices with particle production.

\subsec{Theory II is not Unitary}

The nonunitarity of theory II is a simple consequence of
the fact that an incoming particle-hole pair can dissociate
as in
\fig\dssc{An incoming boson dissociates} into a leftmoving and
rightmoving fermion. One fermion can be reflected back into its
original half space while the other can
tunnel through into the adjoining half space.
The Hilbert spaces for left-
and right- movers are then the one-soliton sectors, which have
no analogue in the Das-Jevicki theory.%
\foot{The Das-Jevicki theory does have soliton sectors
corresponding to one-eigenvalue instantons
\ref\jevsol{A. Jevicki, ``Nonperturbative collective field theory,''
BROWN-HET-807}. It is possible that taking these into account
will restore unitarity within the context of the
collective field theory.}

More formally, we may
consider the unitarity equations for $1\to 2$ scattering on the
same side of the potential. The contributions of
 transmitted amplitudes
to the equation must be of order $e^{-4\pi \mu}$. However, the
reflected amplitudes fail to satisfy the unitarity equations at
order $e^{-2\pi \mu}$, as we now show. Expanding in small
energies $\omega_i$ we have the $S$-matrix elements
\eqn\expdss{\eqalign{
S(\omega,+|\omega,+)={1\over 1+e^{-2\pi\mu}}&\omega+\cdots\cr
S(\omega,+|\omega_1,+;\omega_2,+)&= - {\pi\over 2}
{1\over \cosh^2\pi\mu}\omega_1\omega_2\cr
&+i\omega_1\omega_2(\omega_1+\omega_2)\bigl({1\over 1+e^{-2\pi\mu}}\psi_1
+{\pi\over 2}{1\over \cosh^2\pi\mu}\psi_0\bigr)+\cdots\cr
S(\omega_1,+;\omega_2,+|\omega,+)&=-{\pi\over 2}{1\over \cosh^2\pi\mu}\omega_1
\omega_2 \cr
&+i\omega_1\omega_2(\omega_1+\omega_2)\bigl({1\over 1+e^{-2\pi\mu}}\psi_1
+{\pi\over 2}{1\over \cosh^2\pi\mu}\psi_0\bigr)+\cdots\cr
}}
where $\psi_0 = {\rm Re}(\psi(\half - i\mu))$ and
$\psi_1 = {d \psi_0 \over d\mu}$.
The first term is order $\CO(e^{-2\pi\mu})$ while the second is order 1.
Checking the unitarity equation
$\langle \omega,+|S S^\dagger|\omega_1,+;\omega_2,+\rangle=0$
in a small energy expansion we
see that the $\CO(\omega^3)$
terms, which are perturbative, cancel but
the nonperturbative $\CO(\omega^2)$ terms cannot cancel. Moreover
the contributions of the transmitted amplitudes to the unitarity
equations are $\CO(e^{-4\pi\mu})$. Hence unitarity
is violated at order $e^{-2\pi\mu}$ as expected from the simple physical
picture.

The violation of unitarity in the massless tachyon theory
is extremely suggestive in view of the conjectured
connection of the $c=1$ matrix model to black hole physics
\nref\witten{E. Witten,
``String Theory and Black Holes,'' Phys. Rev. {\bf D44} (1991) 314.}%
\nref\wadseng{G. Mandal, A,M. Sengupta, and S.R. Wadia,
``Classical Solutions of Two-Dimensional String Theory,''
IAS preprint IASSNS-HEP/91/10.}%
\nref\martshat{E. Martinec and S. Shatashvili,
``Black hole  physics and Liouville theory,'' EFI-91-22.}%
\nref\dvv{R. Dijkgraaf, H. Verlinde, and E. Verlinde,
``String propagation in a Black Hole Geometry,''
Princeton preprint PUPT-1252.}%
\refs{\witten-\dvv}.
Unfortunately, that relation must be further
clarified before we can confidently apply the above results
to the black hole problem.

\newsec{Analytic Structure of the $S$-matrix}

\subsec{General Remarks}

In standard field theory the analyticity properties of an
$S$-matrix are related to important physical properties of
the theory.%
\foot{We would like to thank R. Shankar and A. Zamolodchikov
for very useful discussions relevant to this section.}
For example, causality implies analyticity in suitable domains, and
the existence of unstable particles and thresholds imply the existence
of poles and cuts in analytically-continued $S$-matrix elements.
Therefore, in this section we examine the analytic properties
of the $c=1$ $S$-matrix
hoping to understand better the nature of the physics of 2D
strings.

Many aspects of our problem
are different from the more familiar examples of relativistic
field theories in Minkowskian spacetimes. A first difference
is that
in standard $S$-matrix theory
\ref\elpo{R. J. Eden, P. V. Landshoff, D.
Olive and J. C. Polkinghorne, {\it The
 Analytic
 S-matrix}, Cambridge University Press.}\
$S$-matrix elements are considered as analytic functions of
$s_{ijk\dots}=(p_i\pm p_j\pm p_k+\cdots)^2$.  Here we will
analytically continue in the energies $\omega_i$, which
are, roughly speaking, $\sqrt{s_{ijk\dots}}$. A second
(related) difference
is that in our problem we only have time translation and
time reversal invariance so many restrictions of Lorentz
invariance are lost. A third difference is that
the particles in our case are massless, whereas in
more realistic examples massless particles like photons and
gravitons, which
interact with one another in asymptotic regions of spacetime, are
usually excluded from discussion.

The third point above leads to a rather different
analytic structure of $S$-matrix amplitudes
than that normally encountered in massive
relativistic field theory. Consider,
for example, the $2\to 2$ scattering amplitude
in a scalar field theory with particles of mass $m$. In
the complex $s=(p_1+p_2)^2$ plane the amplitude has
the structure indicated in
\fig\stanstrc{Analytic structure in the complex $s$ plane}.
There are elastic threshold branch points at $0,4m^2$. On the
physical sheet there are no other singularities, but
if we analytically continue to the second sheet we may
discover resonance poles at some complex mass-squared
$s=\mu^2$ as indicated in \stanstrc.
Moreover, the presence of these resonance poles implies
the existence of further cuts, for example the one beginning
at $s=(m+\mu)^2$ indicated in \stanstrc. Consider the
limit of \stanstrc\ as $m\to 0$. In this limit the physical
sheet separates into two disjoint half-spaces. There should
be no singularities on the physical sheet and the physical $S$
matrix is defined by two different analytic functions
$S^{\pm}$ defined on these two regions. If we analytically
continue, say, $S^+$ from the upper half plane
into the lower half plane
we will encounter resonance branch cuts located at the position
formerly occupied by the resonance poles.

In the $c=1$ $S$-matrix, the cut structure analogous to the
separation of $S^+$ from $S^-$ above has already been investigated
at tree level in
\ref\kdf{D. Kutasov and Ph. DiFrancesco, ``World Sheet and Space Time
Physics in Two Dimensional (Super) String Theory,''
Princeton preprint PUPT-1276.}.%
\foot{
In \kdf\ D. Kutasov and Ph. Di Francesco observed
an interesting analytic property of
the tree-level $S$-matrix. Namely, by
restricting attention to ``one-particle irreducible''
elements one could work with analytic expressions
(polynomials in the $\omega_i$) valid in all kinematic
regimes. Unfortunately this
property does not persist at higher loops
(as one might guess from physical grounds).
Using the explicit formulae above it is easy to check that the
``one-particle irreducible'' amplitudes defined in \kdf\
are not analytic in the momenta at one loop order.}
We will see below that nonperturbatively
some new singularities arise corresponding to the
presence of resonances.

\subsec{Analytic Properties of Bounce Factors}

The analytic properties of the $S$-matrix follow from
those of the bounce factors. For theories of type I with
$A=0$ the bounce factor is
\eqn\phsfun{\eqalign{
e^{i\Theta(x)}=&
\sqrt{1+i e^{-\pi x}\over
1-i e^{-\pi x}}
\sqrt{\Gamma(\half-ix)\over\Gamma(\half+ix)}\cr
=&\sqrt{2\over\pi}e^{i\pi/4}
\cos\bigl({\pi\over 2}(\half+i x)\bigr)\Gamma(\half-ix)\ .\cr}
}
{}From the first equality it is obviously a phase for $x$
real. From the second we see that it can be continued into
the complex $x$-plane, where it has
simple poles at $x=z_\ell\equiv -i(2\ell+{3\over 2})$ for $\ell=0,1,2\dots$,
with residues $\rho_\ell =\sqrt{2\over\pi}e^{i\pi/4}(-1)^{\ell}{1\over
(2\ell+1)
!}$.
When $A\not=0$ we have instead
\eqn\bncwa{
e^{i\Theta(x,A)}={1+(e^{\pi x}-i)X(x,A)\over
1+(e^{\pi x}+i)X(x,A)}e^{i\Theta(x)
}
}
where
\eqn\trdfx{
X(x,A)\equiv {e^{-\pi x/2}\over \sqrt{2} \pi}
\Gamma({3\over 4}+i{x\over 2})\Gamma({3\over 4}-i{x\over 2})
A { {}_1F_1({3\over 4}-i{x\over 2},{3\over 2};\half i A^2)
\over
{}_1F_1({1\over 4}-i{x\over 2},{1\over 2};\half i A^2) }\ .
}
The bounce factor in \bncwa\ defines a meromorphic function of
$x$ which has a sequence of poles at $x=z_\ell(A)$,
smoothly evolving from the poles at $A=0$. We denote
the residue at the pole by $\rho_\ell(A)$.

The poles $z_\ell(A)$ in the bounce factors
have a simple interpretation
in terms of the free fermions. The denominator of \bncwa\
vanishes for $z$ such that the wavefunction
(see appendix A for a definition) $\chi_R^+(z,A)=0$.
For
$Im z<0$ the wavefunction $\chi_R^+(z,\lambda)$ decays as
$\lambda\to +\infty$ so the condition $\chi_R^+(z,\lambda=A)=0$
is exactly the condition to have a bound state energy
$z$ in the upside down oscillator potential with an
infinite wall at $A$. The ``energy'' $z$ is complex
because this bound state is unstable.
As $A\to -\infty$ there is a deeper
and deeper well on the far side of the parabola and
the resonant state with energy $z_\ell(A)$ becomes more
and more long-lived, i.e., the imaginary parts of $z_\ell(A)$
go to zero. Indeed, the poles $z_\ell(A)$ approach the
positive real axis and become dense there, eventually becoming
a massless free fermion cut.
(It is possible to obtain some analytic results on $z_\ell(A)$
which confirm the above picture deduced on physical grounds.)

\subsec{Theory II and the Limit of Theories of Type I}

One may take the limit $A\to -\infty$ to obtain the bounce factor
of theory $II$ provided one works at $Im z>0$:
\eqn\firstlim{
\lim_{A\to -\infty} R^I_q(z,A)=R^{II}_q(z)\ .
}
At $Im z<0$ this is not true and we have instead
\eqn\seclim{
\lim_{A\to -\infty} R^I_q(z,A)=(1+e^{-2 \pi z})R^{II}_q(z)\ .
}
The reason for the discrepancy is that as $A\to-\infty$ a new cut
for the fermion, the coalescence of the poles
$z_\ell(A)$, appears along the real axis. Indeed, for
$Im z=0$ the $A\to -\infty$ limit of $R^I$ is ill-defined.

The analytic structure of $R^{II}$ may be studied using the formula \gmfrm.
One finds a series of poles in the lower half plane similar to that found
for $R^I$, leading to a similar analytic structure for the scattering
amplitudes in the type-II theory.

\subsec{General Properties of ``Massless Tachyon'' Amplitudes}

We may now continue the amplitudes in \fltfrm\ by choosing
all $\omega_i$ but one to be independent and continuing
in those.
Upon analytic continuation the integrals in \fltfrm\ become
contour integrals, and singularities arise when the endpoints
of the integrals hit poles of the integrand or when poles of the
integrand pinch the contour. Since we have written the integral
as a sum of terms there will in general be several choices for how
to continue the contours in the separate integrals. One can
define a notion of a ``physical sheet'' by starting in a given
kinematic region and analytically continuing into an appropriate
half-space. Note that if a bounce line carries positive
energy $\omega$ then the corresponding factor
$e^{\pm i\Theta(\mu\pm \omega)}$ has poles
at $\omega=\mp \mu-i(2\ell+{3\over 2})$.
Therefore, from rule RH1, if we define the physical sheet to
be included in the half-space
$Im(\omega)>0$, there can be no singularities in the integral.
On the other hand, it can happen that when we continue
out of this half-space a pole in the integrand
is forced to hit the endpoint of the contour integral.
This results in a branch cut singularity. Moreover special
configurations of $\omega_i$ can produce more complicated
singularities, and continuation to further sheets can produce
new singularities. Such singularities are indicative of
new states and degrees of freedom in the system not directly seen
in the physical $S$-matrix. These points are best illustrated by
explicit examples.

\subsec{Analytic Structure of $S(\omega|\omega)$}

In the $1\to 1$ amplitude
\eqn\essoo{
S(\omega|\omega;A)=
\int_0^{\omega} d x
e^{i\Theta(\omega-\mu-x,A)
+i\Theta(\mu+x,A)}
}
the integrand has a set of fixed poles at $x=-\mu+z_\ell(A)$ and
set of moving (with $\omega$) poles at $x=\omega-\mu-z_\ell(A)$. The contour
integral is therefore forced to hit these poles when
$\omega=\pm \mu + z_\ell(A)$ and at these points the integral
has a logarithmic singularity. We define the second sheet by
choosing the branch cuts as in
\fig\cutplane{A set of branch cuts for $S(\omega|\omega)$, illustrated
here fore the case $A=0$.}.
{}From the discussion of section 5.2 we see that the branch cut
singularities are naturally described by a simple physical picture:
one fermion gets
trapped in a metastable state. The other fermion bounces off
the potential and can radiate and reabsorb
massless tachyons, thus producing a cut singularity.

It is interesting to proceed and continue to the third sheet by
passing through one of the cuts of \cutplane. For simplicity
we consider continuing through a cut at $\omega=-\mu+z_\ell(A)$,
although a similar story holds for the cuts at $\omega=\mu+z_\ell(A)$.
Comparing the contour
integrals defined in the $x$ plane in
\fig\twopaths{Two paths defining the $1\to 1$ amplitude on the
second and third sheets.}\ we see these differ by
\eqn\thrdsec{
S^{(3)}(\omega|\omega;A)=S^{(2)}(\omega|\omega;A)+\rho_\ell(A)
e^{i\Theta(\omega
-z_\ell(A),A)}
}
where the superscript refers to sheet number.
The extra factor produces a new set of poles on the third sheet
at $\omega=z_\ell(A)+z_{\ell'}(A)$. These poles may be interpreted as
a metastable state obtained when a particle-hole pair
resonates. Since the fermions are free the ``energies'' $z_\ell$ simply add.
Since both are trapped, neither fermion can shake off massless
tachyons so the singularity produces a pole and not a cut.
Note that when $A=0$ these poles occur at $\omega=-in$ for
$n=3,5,\dots$.

\subsec{Analytic Structure of the Four-point Function}

Proceeding along similar lines one can investigate in detail,
e.g., the four-point function $S(\omega_1,\omega_2|\omega_3,\omega_4;A)$
for $\omega_2<\omega_3,\omega_4<\omega_1$. We take $\omega_1$ to
be dependent and continue from a real subspace of $\IR^3$.
One finds a rather complicated singularity structure. The most
interesting singularities appear to be double-poles arising from
the four-bounce terms. These correspond to processes where an
incoming tachyon (say, $|\omega_2\rangle$) becomes trapped in a
resonant state $\omega_2=z_\ell(A)+z_{\ell'}(A)$ which later decays
to $\langle \omega_4|$. More precisely, letting
$\omega_2=z_\ell + z_{\ell'}+\epsilon_1$ and $\omega_4=z_\ell +
z_{\ell'}+\epsilon_2$
with the $\epsilon$'s small we get double poles
\eqn\dbleples{\rho_\ell \rho_{\ell'}^2 \biggl[{1\over
\epsilon_1 \epsilon_2}e^{i
 \Theta(\omega_3-z_\ell)} +{1\over \epsilon_1
(\epsilon_1-\epsilon_2)}e^{i \Theta(\omega_3-z_{\ell'})}\biggr]}
The residue of this double pole, which is
essentially a bounce factor may be interpreted as an amplitude for
scattering off an ``excited background''
corresponding to adding a single resonant tachyon to the fermi sea.

\subsec{Lessons from the Analytic Structure of $S$}

The main lesson we may draw from these remarks is that it is
not possible to neglect the fermionic degrees of freedom
beyond perturbation theory. Even in theories of type I, where
we have a nonperturbatively
unitary $S$-matrix without including soliton sectors or matrix
model fermions,
the existence of the fermions can be detected by studying the
analytic properties of the $S$-matrix to find resonance poles
and cuts. Presumably the nature of the residues and discontinuities
at these singularities would indicate the fermionic
nature of the particles. It must be emphasized that these
analytic singularities are not mathematical artifacts but have
real effects. If $A$ is large and negative the resonant
states we have discussed have long lifetimes. If, for example,
these lifetimes exceed those of experimentalists measuring
the $S$-matrix at $\lambda=+\infty$ they will find themselves
puzzling over an apparent loss of unitarity.

A second lesson is that the rich singularity structure
of the higher point amplitudes indicates the existence of
a correspondingly rich spectroscopy of excited, but unstable,
backgrounds. These time-dependent backgrounds, resulting from
changes in the Fermi sea, are the nonperturbative analogues of
the tachyonic backgrounds studied by Polchinski in
\joefluid.

\newsec{Worldsheet Interpretation}

\subsec{Connection to Liouville Theory}

So far our discussion has emphasized the free-fermion
formulation of the matrix model. Since the double-scaled
matrix model is a sum over continuum surfaces we expect
that the continuum amplitudes can also be described by
the conformal field theory of a massless scalar $X$
coupled to a $c=25$ Liouville theory $\phi$ with worldsheet
action
\eqn\confld{
\CA=\int \half\p X\pb X + \p\phi\pb\phi +\sqrt{2} R^{(2)}\phi
+\mu e^{\sqrt{2}\phi}
}
As explained in \lpsflds\ the Liouville and $\tau$
coordinates are not the same. Recall that the objects calculated
in a sum over continuous geometries on 2-surfaces with
boundary are ``macroscopic loop
amplitudes'' defined by fixing the boundary values of the
two-metric $e^{\sqrt{2}\phi}$ so that the bounding
circles $\CC$ have lengths $\ell=\oint_\CC e^{\phi/\sqrt{2}}$
\nref\bdss{T. Banks, M. Douglas, N. Seiberg, and S. Shenker, ``Macroscopic
and Microscopic Loops in Nonperturbative Two-Dimensional Quantum
Gravity,'' Phys. Lett. {\bf 238B} (1990) 279.}%
\nref\ambjorn{J. Ambjorn, J. Jurkiewicz, and Yu. Makeenko, ``Multiloop
Correlators for two-dimensional Quantum Gravity,'' Phys. Lett.
{\bf 251B} (1990) 517.}%
\nref\kostov{I.K. Kostov, ``Loop Amplitudes for Non Rational String
Theories,'' Phys. Lett. {\bf 266B} (1991) 42,317.}%
\nref\mms{E. Martinec, G. Moore, and N. Seiberg,
``Boundary Operators in 2D Gravity,'' Phys. Lett. {\bf 263B} 190.}%
\nref\lpstates{G. Moore, N. Seiberg, and M. Staudacher, ``From Loops
to States in Two-Dimensional Quantum Gravity,''
Nucl. Phys. {\bf B362} (1991)}%
\refs{\bdss--\lpstates}.

In
\refs{\moore,\lpstates}
it was proposed that one extract the $c=1$ correlators of tachyons
by extracting the terms proportional to nonanalytic powers
$\ell^{|p|}$ in the small $\ell$ expansion of the macroscopic loop amplitudes.
{}From the continuum 2D path integral point of view $W(\ell,p)$
corresponds to an expansion of a sum of local operators.
As discussed in \lpsflds\ that sum of operators can be written as
\eqn\locopexp{W_{in}(\ell,p)= -\CT_p {\pi\over sin\pi |p|} \mu^{-|p|/2}
I_{|p|}(2\sqrt{\mu}\ell)-
\sum_{r=1}^\infty\hat\CB_{r,p}{2(-1)^r r\over r^2-p^2}
\mu^{-r/2}I_r(2\sqrt{\mu}\ell) }
where $\CT_p$ is proportional to the tachyon vertex operator and
$\hat\CB_{r,p}$ are redundant operators for $p\notin \IZ$.

We now relate the prescription of this paper to that of
\refs{\moore,\lpstates}.
{}From the matrix-model representation of macroscopic loop
operators $W(\ell,q)$ we see
that the amplitudes are related by an integral transform:
\eqn\mclp{
\langle \prod W(\ell_i,q_i)\rangle_c\equiv\int_A^\infty \prod d\lambda_i
e^{- \ell_i\lambda_i}\langle\prod \rho(\lambda_i,q_i)\rangle_c
}
in theories of type I.%
\foot{In theories of type II we must be more careful since the eigenvalue
distribution grows on both sides of the axis. The procedure
advocated in \moore\ is to take a Fourier transform with
$\ell=i z$, $z$ real, and analytically continue positive and
negative frequency components separately.
This procedure is correct to all orders
of perturbation theory, but its status as a nonperturbative
definition is unclear. Hence we limit our discussion to
theories of type I.}
The $\ell\to 0$ asymptotic behavior of the integral \mclp\  is
dominated by the $\lambda\to\infty$ behavior. Changing variables to
$\tau$  and using the asymptotic expression \defrn\ we see that
\eqn\mclpasy{
\langle\prod w(\ell_i,q_i)\rangle\quad
{\buildrel \ell\to 0\over \sim}\quad \int^\infty \prod d\tau_i
e^{- \ell_i2\sqrt{\mu}\cosh\tau_i}\prod e^{-|q_i|\tau_i}
\biggl[\CR_n+\CO(e^{-\tau_i})\biggr]
}
For small $z$ and $q\notin\IZ$ we have the asymptotics:
\eqn\intgl{\int^\infty d\tau
e^{-\ell 2\sqrt{\mu}\cosh\tau}e^{-|q|\tau}\sim
(\ell\sqrt{\mu})^{|q|}\Gamma(-|q|)
}
plus terms regular in $\ell$. Hence
$\langle \prod \CT_{q_i}\rangle=\CR(q_1,\dots, q_n;\mu)$.
To make a connection to standard vertex operator normalizations we
compare with the vertex operator calculations found, e.g., in
\kdf\ which show that
\eqn\vertnorm{
\CT_q={\Gamma(|q|)\over \Gamma(-|q|)}\int_\Sigma e^{iq X/\sqrt{2}}
e^{\sqrt{2}(1-\half |q|)\phi}\equiv {\Gamma(|q|)\over \Gamma(-|q|)} V_q
}
Thus we finally have a prediction for all
integrated vertex operator correlators
in the $c=1$ theory:
\eqn\modspace{\prod_i {\Gamma(-|q_i|)\over \Gamma(|q_i|)}\CR
_n(q_1,\dots,q_n;\mu)\quad
{\buildrel \mu\to \infty\over \sim}\quad
\sum_{g\geq 0} \biggl({1\over \mu}\biggr)^{2g-2+n}
\int_{\CM_{g,n}}
\langle \prod_{1}^{3g-3+n} b\bar{b} \prod_{i=1}^n
c\bar{c}V_{q_i}\rangle\biggr|_{\mu=1}
}
where $\CM_{g,n}$ is the moduli space of curves of genus $g$ with
$n$ punctures.

(Let us note parenthetically that
the ``wavefunction normalization'' factors
$ {\Gamma(-|q_i|)\over \Gamma(|q_i|)}$
in \modspace\ have been the subject of much
discussion
\nref\polyi{A.M. Polyakov, ``Self-Tuning Fields and  Resonant Correlations
in 2D-Gravity,'' Mod. Phys. Lett. {\bf A6} (1991) 635.}%
\nref\grosetal{D.J. Gross, I.R. Klebanov, and M.J. Newman,
``The Two-Point Correlation Functions of the One Dimensional Matrix Model,''
Nucl. Phys. {\bf B350} (1991) 621 \semi
D.J. Gross and I.R. Klebanov, ``$S = 1$ for $c = 1$,'' Princeton preprint
PUPT-1241 \semi
D.J. Gross and U.H. Danielsson, ``On the Correlation Functions of the
Special Operators in $c=1$ Quantum Gravity,'' Princeton preprint
PUPT-1258.}%
\nref\minic{D. Minic and Z.Yang, `` Is $S=1$ for $c=1$?''
Texas preprint UTTG-23-91.}%
\nref\tanii{N. Sakai and Y. Tanii, ``Factorisation and Topological states
in $c=1$ Matter Coupled to 2D Gravity'' TIT/HEP-173.}%
\refs{\polyi-\tanii,\kdf}.
These factors are singular when $q\in \IZ$ and it has ben proposed that the
external line divergences should be reinterpreted as the contributions of
``intermediate'' on-shell states (in some space-time field theory).
Such an interpretation requires inclusion of the
``special state'' vertex operators obtained from dressing
nontrivial Virasoro primaries in Fock spaces with
charge $p\in \IZ$ with highest Virasoro weight
${n^2\over 4}$. An alternative
interpretation of these factors was proposed in \lpsflds\
based on the smoothness of macroscopic loops for $q\in\IZ$.
The apparent singularity arises since the extraction of
nonanalytic powers of $\ell$ is not well-defined for
$q\in\IZ$. Indeed the integral \intgl\ can be written
more precisely as
\eqn\carinte{\eqalign{
\int_A^\infty d\tau e^{-2\ell\sqrt{\mu}\cosh\tau}e^{-|q|\tau}=&
-{\pi\over \sin\pi|q|}I_{|q|}(2\sqrt{\mu}\ell)\cr
-\sum_{n\geq 0}
(-1)^n &(\ell\sqrt{\mu})^n\sum_{m=0}^n{1\over m!(n-m)!}
{e^{A(m-n-|q|)}\over m-n-|q|}\cr}
}
Clearly the left hand side is smooth as $q$ becomes integral, but
the decomposition of the right hand side breaks down.
This was interpreted in \lpsflds\ as the fact
that $\CB_{s,q}$ while in general a redundant boundary operator
becomes a bulk operator at special values of $q$. It thus appears to
be consistent to use the vertex operator normalization
$\CT_q$, bearing in mind that for $q\in\IZ$ the operator
is a linear combination of special state and tachyon vertex
operators.)

\subsec{Continuations to Minkowski Space}

The $c=1$ $\times$ Liouville system is a conformal field
theory with two Euclidean signature bosons. The Minkowskian
spacetime physics deduced from this Euclidean theory
depends strongly on how we analytically continue.

The analytic continuation described in section 2 above
identifies $X$ as Euclidean time. Thus we continue
$X\to i t$ and $|k|\to -i \omega$. If we formally
continue the expression in \modspace\ we obtain
correlators of macroscopic state operators
\ref\natiliouv{N. Seiberg, ``Notes on Quantum liouville Theory and
Quantum Gravity,''
in
{\it Common Trends in
Mathematics and Quantum Field Theories,} Proceedings of the 1990 Yukawa
International Seminar, Prog. Theor. Phys. Supp. {\bf 102}.}
\eqn\mcstate{
\int e^{\pm i\omega t(z,\bar z)} e^{(\sqrt{2}+i\omega)\phi(z,\bar z)}\qquad .}
These operators have positive Liouville energy and, at least
semiclassically, create surfaces with large holes and singular
metrics. Indeed the difficulties associated
with the so-called ``$c=1$ barrier'' were ascribed by N. Seiberg
to the
destruction of a smooth world sheet arising from
inclusion of such operators in the action \natiliouv.%
\foot{See
\ref\joetalk{
J. Polchinski, ``Remarks on the Liouville Field Theory,''
UTTG-19-90, to appear in Strings '90, Texas AM.} for a different
opinion.}
It would therefore be very interesting
to see if the formal analytic continuation of the RHS of
\modspace\ can be given a more rigorous justification.

The analytic continuation of $X$ to Minkowskian time
also leads to a simple spacetime interpretation of the
Seiberg bound \refs{\natiliouv,\joetalk} which states that
Liouville exponentials $e^{\alpha\phi}$ must satisfy
$\alpha<\half Q$. The fact that only one root of the KPZ
equation is allowed is simply the fact that
for scattering on the right half-line, incoming particles
must be leftmovers and outgoing particles must be rightmovers.
Note in this connection that if we work in theory II, with
another asymptotic universe then, nonperturbatively, there
can exist ``wrong branch'' states, i.e., incoming rightmovers
from the point of view of the righthand universe.%
\foot{A related remark was made by N. Seiberg. He noted
that the wavefunctions of \lpstates\ associated to the
``wrong branch''  could be obtained from continuations
appropriate to the far side of the parabola.}

Finally the issue of correct normalization of tachyon
vertex operators discussed above becomes academic in
this continuation \klebrev. The $S$-matrix obtained
by continuing $\CR_n$ is simply related by the
phase redefinition of statevectors
\eqn\rrenv{\eqalign{
|\omega_1\dots \omega_n\rangle &\to
\prod{\Gamma(i\omega_i)\over\Gamma(-i\omega_i)}
|\omega_1\dots \omega_n\rangle\cr
\langle\omega_1\dots \omega_n| &\to
\prod{\Gamma(i\omega_i)\over\Gamma(-i\omega_i)}
\langle \omega_1\dots \omega_n|\cr}
}
to the $S$-matrix obtained from vertex operator calculations.
Hence, no probability amplitudes are changed and the theories
are physically indistinguishable.

Another choice of continuation which is sometimes adopted is to regard
the Liouville coordinate as Euclidean time. This prescription
``$\phi \to i \phi$'' is rather less well-defined. Indeed
serious objections to it have been raised in \joetalk.
Roughly speaking time becomes effectively semi-infinite,
the Liouville wall corresponding to something like a big bang
or a big crunch. In the former case the amplitudes
$\CR_n(q_1,\dots ,q_n)$
correspond to ``amplitudes'' for producing right-movers and
leftmovers (corresponding to positive and negative $q_i$)
from the finite past. Since time is semi-infinite there is no
obvious analog to the unitarity constraint.

\subsec{Mathematical Applications}

The equality of the asymptotic expansions in \modspace\ identifies
the coefficients of the large $\mu$ asymptotics of
$\CR_n(p_1,\dots ,p_n;\mu)$ with integrals
over $\CM_{g,n}$ of some natural densities provided by the
Liouville theory. From our construction of
$\CR$ the answers are simply expressed in terms of Bernoulli
numbers, thus giving formulae reminiscent of
\nref\wittgauge{E. Witten, ``On Quantum Gauge Theories in Two
Dimensions,''  IASSNS-HEP-91/3}%
\nref\penner{C. Itzykson and J. B. Zuber,
``Matrix Integration and Combinatorics of Modular Groups,''
Comm. Math. Phys. {\bf 134} 197 (1990) \semi
J. Distler and C. Vafa,
``The Penner Model and $D=1$ String Theory,''
Princeton preprint PUPT-1212, lecture
given at Carg\'ese Workshop on Random Surfaces,
Quantum Gravity and String Theory, 1990}%
\refs{\wittgauge,\penner}.
These observations add more
evidence to the oft-quoted conjecture that there is a
topological field theory version of the $c=1$ model.
To make the connection precise one should (1) generalize our
formulae to the self-dual radius  and (2) understand the appropriate
{\it nontrivial} $k\to \infty$ limit of the twisted $N=2$ minimal model.
Step (1) might be accomplished
using the results of
\nref\grocirc{D.J. Gross and I.R. Klebanov,
``One Dimensional String Theory on a Circle,''
Nucl. Phys. {\bf B344} (1990) 475.}%
\nref\klelow{I. Klebanov and D. Lowe,
``Correlation functions in 2D Quantum Gravity Coupled
to a Compact Scalar Field,'' Princeton preprint PUPT-1256'' }%
\refs{\grocirc,\klelow},
but step (2) appears to be more
difficult.

\newsec{Background Dependence of the S-matrix}

The most interesting questions in 2D string
theory center on the nature of strings in different
backgrounds. It should be possible to perturb the
conformal field theory \confld\ to obtain more general
nonlinear sigma models with two target space
coordinates $(X,\phi)$. In general, nearby theories
are expected to have an action of the schematic form
\eqn\action{
\CA+\delta \CA=
\int_\Sigma \biggl[\half \p X\pb X + \p\phi\pb\phi +\sqrt{2} R^{(2)}\phi
+\mu e^{\sqrt{2}\phi}\biggr] +\int dp \epsilon(p)V_p +
\sum \hat t_{r,s} V_{r,s}
}
where $V_{r,s}$ denotes some basis of ``special state operators.''

As long as the background perturbations preserve conformal invariance and
the asymptotically flat nature of the target space geometry, the string
$S$-matrix should make sense, and we should therefore discuss the
functions
$S(\{\omega_i\} \to \{\omega_i'\};{\rm backgrounds})$
on the infinite-dimensional space of background geometries.
These can be parametrized, in an infinitesimal neighborhood
of \confld\ by a subspace spanned by the directions
$\epsilon(p), \hat t_{r,s}$. Ultimately, we would
like to know what principle determines the allowed backgrounds
and what equations govern the background dependence of $S$.

The background dependence on $\epsilon(p)$ is an interesting
question which we will address elsewhere.
For the present we describe how $S$ changes in an
infinitesimal neighborhood around \confld\ by perturbing
the parameters $\hat t_{r,s}$. The matrix-model
identification
\eqn\vtobe{
V_{r,s}\leftrightarrow \hat\CB_{s,q=r}=\int \lambda^s e^{ir x}
\psi^\dagger\psi(\lambda,x) d\lambda dx + \cdots
}
for $r\in\{s,s-2,\dots,-s\}$ suggests, by exponentiation, that
perturbations of the $\sigma$-model action by the parameters
$t_{r,s}$ correspond to perturbations of the matrix model
potential
$V(\lambda)=-\half \lambda^2+\sum {t}_{r,s}\int
\lambda^s e^{irx}$.%
\foot{The meaning of the ellipsis in
\vtobe\ and the hat notation
is explained in \refs{\lpstates,\lpsflds}.}
Moreover, this reasoning suggests that changes in $V$ which
affect the nonperturbative $S$-matrix but not its perturbation
expansion could be interpreted as ``small'' perturbations in the
background geometry which only affect nonperturbative physics.
The dependence on $A$ discussed in section five above is an example
of such a perturbation.

The change in the $S$ matrix due to a change in background
corresponding to $\delta V(\lambda)$ is easily computed (in principle)
since the fermions remain free, and one need only compute
the change in the fermion two-point function. In particular,
for time-independent perturbations we need only
compute the change in the resolvent \prev:
\eqn\chgeres{
I\to \langle \lambda_1| {1\over H+\delta V-z}|\lambda_2\rangle
}
To be more explicit we must realize that the operators
$\int \lambda^s$ are ill-defined due to ultraviolet
worldsheet divergences
\refs{\moore,\lpsflds}
so we consider matrix-model
potentials $\delta V(\lambda)$ with rapid falloff for
$\lambda\to \infty$. These may be viewed as linear
combinations of special state operators.
The new $S$-matrix is obtained from
the old by a simple change of the bounce factor which is given
to lowest order by (we put $A=0$ here):
\eqn\chgebnce{
\delta\Theta(z)=-\pi\int_0^\infty (\psi^-(z,\lambda))^2\delta V(\lambda)
d\lambda +\cdots
}

The background dependence of the scattering matrix leads to
background dependence of
the location of the singularities described in section five.
The variation with background can be computed in the $A=0$ theory
using a trick. The poles $z_\ell$ are bound state poles, and
indeed at $z_\ell=-i(2\ell + {3\over 2})$ the bound state
wavefunctions are just given by the continuation of harmonic
oscillator wavefunctions:
\eqn\bdstwv{
\psi_\ell(\lambda)=
H_{2\ell+1}(2^{-1/2}e^{i\pi/4}\lambda)e^{-i\lambda^2/4}
}
which satisfy the identity:
\eqn\orthnr{\int_0^\infty \psi_\ell(\lambda) \psi_{\ell'}(\lambda) d\lambda
= 2^{2\ell+\half}\sqrt{\pi}(2\ell+1)!e^{-i\pi/4}\delta_{\ell,\ell'}
}
and are thus orthogonal {\it without} complex conjugation.
It follows that if $\delta V(\lambda)$ has an expansion in terms of even
powers of $\lambda$ (these correspond to the nonredundant operators
\refs{\wvii,\lpsflds})
then the variation of the bound-state pole with $\delta V$ is given
by the expectation value in the normal harmonic oscillator:
\eqn\chgeofz{
\delta z_\ell=\langle 2\ell+1|\delta
V(\sqrt{2}e^{-i\pi/4}x)|2\ell+1\rangle
}
where $|n\rangle$ is an orthonormal basis for the standard
harmonic oscillator.
Thus the locations of the singularities $z_\ell$ change as we
``turn on'' the special state operators displaying the
background dependence of these singularities.

\newsec{2D String Physics at High Energies}

The importance of studying string theory in the limit of
ultrahigh energies has been emphasized by several authors
\nref\grmend{D.J. Gross and P.F. Mende,
``The High Energy Behaviour of String Scattering Amplitudes,''
Phys. Lett. {\bf 197B} (1987) 129;
``String Theory beyond the Planck Scale,''
Nucl. Phys. {\bf B303} (1988) 407.}%
\nref\venez{D. Amati, M. Ciafaloni, and G. Veneziano,
``Superstring Collisions at Planckian Energies,''
Phys. Lett. {\bf 197B} (1987) 81.}%
\nref\grspec{D. Gross,   ``High Energy Symmetries of String Theory,''
Phys. Rev. Lett. {\bf 60} (1988) 1229.}%
\nref\witspec{E. Witten, ``Space-Time and Topological Orbifolds,''
Phys. Rev. Lett. {\bf 61} (1988) 670.}%
\nref\menoog{P.F. Mende and H. Ooguri,
``Borel Summation of String Theory for Planck Scale Scattering,''
Nucl. Phys. {\bf B339} (1990) 641.}%
\refs{\grmend-\menoog}.
In this section we will throw some light -- literally
-- on the
nature of the tachyon condensate wall by studying the
reaction of the wall long after a single boson of energy $\omega$
has been thrown at it. It was pointed out in \moore\ that
at large energies $\omega\gg \mu$, with fixed string coupling, the asymptotic
expansions of string perturbation theory cease to make sense and
there must be new physics in this regime. We emphasize that this large order
behavior of perturbation theory is in principle derivable from the
continuum Liouville approach and therefore a true string-theoretic
effect, independent of the matrix model regularization. On the
other hand,
the particular kind of new physics that arises is probably sensitive
to the nonperturbative definition of the model.
In what follows we will attempt to study this new physics concentrating on
the nonperturbative definition we have labeled
theory I with $A=0$.

\subsec{Particle Production}

The unitarity equations for $1\to 1$ scattering read:
\eqn\unitonon{
1={1\over \omega^2}|S(\omega|\omega)|^2
+{1\over \omega}\sum_{n\geq 2}
\int_0^\infty\prod_{i=1}^n{d \omega_i\over \omega_i}
\delta(\sum_{i=1}^n \omega_i-\omega)|S(\omega|\omega_i)|^2
}
Thus, the probability that a single incoming boson of energy $\omega$
produce two or more outgoing particles is simply expressed as
\eqn\partprod
{P(\omega)=1-\Biggl|\int_0^1 dt
e^{i\Theta(t\omega+\mu)
+i\Theta((1-t)\omega-\mu)}\Biggr|^2
}
At large $\omega$ this behaves like $P(\omega)\sim 1-{\pi\over 2\omega}$.
Nevertheless there is an intricate resonant structure in $P(\omega)$
for finite $\omega$.
We have plotted this function in
\fig\pltprtprd{The amplitude $1-P(\omega)$ for $\mu=8$
as a function of $\omega$.
The complicated structure observed here is probably not an
artifact of the existence of the wall. Indeed we may also plot
the amplitudes in theory II to obtain similar pictures.}

\subsec{Energy Distributions}

We can also look explicitly at the probability that two
particles of energies $\omega_1+\omega_2=\omega$ are created.
The result can be described in terms of a normalized energy
distribution for $\omega_2$
\eqn\onetwo{D(\omega_2,\omega;\mu)=
\half {1\over (\omega -\omega_2)\omega_2}|\CA|^2}
where
\eqn\ampint{\eqalign{
\CA=&T(\omega_2;\omega,-\mu)-T(\omega_2;\omega,\mu)\cr
&=\int_{-\mu}^{\omega_2-\mu}dx
e^{i\Theta(\omega-x)+i\Theta(x)}-
\int_{\mu}^
{\omega_2+\mu}dx
e^{i\Theta(\omega-x)+i\Theta(x)}\cr}
}
Again this distribution has very interesting large-energy behavior.
One finds that the phase integral defining
$T(x;\omega,\mu)$ has no stationary phase point
for $x<\half \omega-\mu$, the integral being of order $1/log \omega$.
At $x\cong \half \omega-\mu$ the integral turns on rapidly  and is
of order $(\pi \omega/2)^{1/2}$. Thus we may approximate $T$ by a
step function. For
$\half \omega-\mu<\omega_2<\half\omega+\mu$,
$\CA$ is flat and of order $(\pi \omega/2)^{1/2}$.

Outside of this range one may approximate
(here we take $\mu\ll \omega_2\ll \omega$):
\eqn\appxa{
\CA\sim {sin(\Theta(\mu)+\mu log\omega)\over log\omega}\Biggl[1-
e^{i\omega_2 log\omega} e^{i\Theta(\omega_2)}\Biggr]
}
up to some irrelevant phases.
The rapidly oscillating second term leads to resonant behavior.
A typical energy distribution is shown in
\fig\enerdis{Energy distribution integral $|\CA|^2$ for $\mu=5$
and $\omega=6$. Again, similar results hold in theory II.}

The above analysis leads to a simple approximate expression for the
matrix element for particle creation
$\omega\to \omega_1+\cdots+\omega_n$
in the case where no $\omega_i$ is small:
\eqn\lrgeng{S\cong e^{2i\Theta(\half \omega)}
\sqrt{i\pi \omega\over 2}\sum_{S\subset S^+}
\theta(\omega(S)-\half \omega+\mu)(-1)^{|S|}
}
Note that this is much softer than the behavior obtained at any order
of perturbation theory.

\subsec{Particle Number Distributions}

At low energies and weak string coupling it is easy to estimate
the particle number distribution defined by the $n^{th}$
term in the series \unitonon :
\eqn\prtnbds{
P(n,\omega)\sim \mu^2 \biggl({
\omega\over \mu}\biggr)^{2n} {((n-2)!)^2\over n!(2n-1)!}
\sim \mu^2 \biggl({\omega\over 2\mu}\biggr)^{2n} {1\over n!n^{5/2}}
}
This is almost a Poisson distribution: at low energies the fermions
bounce with small phase shifts, and the subsequent bosonization proceeds
via independent processes. The high-energy distribution function
would be extremely interesting to work out. It is rather difficult
since the approximation \lrgeng\ breaks down near the boundaries
of the simplex $\sum \omega_i=\omega$ and these boundaries contribute
significantly to the answer. We must leave the analysis of this point
for the future.

\newsec{Conclusions}

Finally, we discuss some points of possible relevance to future work.

The nonperturbative violation of
unitarity in theory II deserves to be understood
better. From the discussion in sec.~4.2 it is clear
that theory II can be made unitary by the inclusion of all the soliton
sectors. This suggests the following
interesting open question. Let us begin with theory II and
consider the most general way of adding degrees of freedom
to restore unitarity. Will we be uniquely led to the free fermions of the
matrix model? It is natural to wonder if a similar situation will some day
occur in higher dimensional string theories, and that nonperturbative
unitarity will unveil the correct degrees of freedom with which
we should describe the theory.

We have made some preliminary remarks on backgrounds,
the main point being that all background dependence on
$V(\lambda)$ can be summarized in the bounce factor
$e^{i \Theta(x;V)}$, which
is essentially the scattering matrix of a free fermion
in the potential $V$.
It is thus clear that
inverse scattering theory is well-suited to this problem, and
we hope to investigate this point more thoroughly in the future.%
\foot{In this connection
C. Vafa has made the interesting suggestion that the
bounce factor be regarded as the $KP$ tau
function and that it should satisfy some difference equations
analogous to the string equation.
One could then guess that background dependence
would essentially be KP flow on the bounce factor.}

Several other points must be clarified before we can achieve a
complete understanding of background dependence.
In general, the relation between
the Liouville coordinate and the matrix model eigenvalue should be
better understood. Also, the important underlying $W_\infty$ symmetry
of the $c=1$ theory has played a minor role in this paper.
We expect that a complete understanding of the background
dependence will require some nontrivial use of this symmetry.
That in turn will require a better understanding of how $W_\infty$ acts
on $S$ than was obtained in section seven.

We have discovered nonperturbative dependence on parameters
like the position of the infinite wall at $\lambda = A$.
Recall that in double-scaling the theory the wall is
at a distance of order $1/N$ from the quadratic maximum so we
are modifying the potential in the scaling region and cannot really
expect physics to be independent of $A$. Thus, we should
not pessimistically call the $A$-dependence a nonperturbative
violation of universality, rather, we should optimistically term
it the discovery of nonperturbative parameters analogous to
the theta parameter of QCD.%
\foot{History repeats itself here:
see the conclusions to the paper of Douglas and Shenker
\dgsh.}
The existence of such parameters, and indeed the existence
of different ``consistent'' backgrounds raises once again
the persistent spectre of the unpredictability of string theory.
There are two possible ways this situation can be remedied.
First, there might be further consistency conditions
not yet considered which might rule out some
possibilities. The absence-of-real-poles constraint on
solutions to Painlev\'e I is a case in point.
In our case we have seen that the $S$-matrix has a very
intricate analytic structure. Accordingly there might be physically
motivated constraints on this singularity structure which
exclude some backgrounds from consideration. Second,
some dynamical principle might favor some backgrounds
and exclude others. Of course, this is an ancient hope,
but with the advent of solvable string models and a deeper
understanding of backgrounds an answer might be forthcoming.

\bigskip
\centerline{\bf Acknowledgements}

We would like to thank L. Susskind for asking a good question.
We would also like to thank
T. Banks, I. Cherednik, L. Crane, D. Kutasov, H. Saleur,
N. Seiberg, R. Shankar, S. Shenker,  C. Vafa,
A. Zamolodchikov, and G. Zuckerman for very helpful discussions and
correspondence. We are also grateful to N. Seiberg and S. Shenker for
comments on a previous draft of the manuscript.
This work is supported by DOE grant DE-AC02-76ER03075
and by a Presidential Young Investigator Award.

\appendix{A}{The Resolvent of the Upside-down Oscillator}

We list here some relevant formulae for working out the large
$\lambda$ asymptotics of
$I(q,\lambda_1,\lambda_2)=(I(-q,\lambda_1,\lambda_2))^*=i\langle\lambda_1
|{1\over H-\mu-iq}|\lambda_2\rangle$ defined in \iasfg.

This resolvent can itself be expressed in terms of parabolic cylinder
functions at a complex energy $z=\mu+i q$. The asymptotics of the
even and
odd parabolic cylinder functions, defined in terms of degenerate
hypergeometric functions by:
\eqn\wvfnspl{\eqalign{
\psi^+(a,x)&={1\over \sqrt{4 \pi (1+e^{2\pi
a})^{1/2}}}(W(a,x)+W(a,-x))\cr
&={1\over \sqrt{4 \pi (1+e^{2\pi a})^{1/2}}}2^{1/4}\biggl|{\Gamma(1/4+i
a/2)
\over \Gamma(3/4+i a/2)}\biggr|^{1/2}e^{-i x^2/4}
{}_1F_1(1/4-ia/2;1/2;i x^2/2)\cr
&={e^{-i \pi/8}\over 2\pi} e^{-a \pi/4}|\Gamma(1/4+i
a/2)|{1\over \sqrt{|x|}}
M_{ia/2,-1/4}(i x^2/2)\cr}
}
\eqn\wvfnsmi{\eqalign{
\psi^-(
a,x)&={1\over \sqrt{4 \pi (1+e^{2\pi a})^{1/2}}}(W(a,x)-W(a,-x))\cr
&={1\over \sqrt{4 \pi (1+e^{2\pi a})^{1/2}}}2^{3/4}\biggl|{\Gamma(3/4+i
a/2)
\over \Gamma(1/4+i a/2)}\biggr|^{1/2} x e^{-i x^2/4}
{}_1F_1(3/4-ia/2;3/2;i x^2/2)\cr
&={e^{-3 i \pi/8}\over \pi} e^{-a \pi/4}|\Gamma(3/4+i a/2)|
{x\over |x|^{3/2}}
M_{ia/2,1/4}(i x^2/2)\cr}
}
are most simply expressed in terms of left and right-mover ``plane wave''
combinations:
\eqn\planewvii{
\eqalign{
\chi^{\pm}_R&\equiv \bigl(\sqrt{k}\mp {i\over \sqrt{k}}\bigr)\psi^+
-\bigl(\sqrt{k}\pm {i\over \sqrt{k}}\bigr)\psi^-
{\buildrel \lambda\to +\infty\over\sim}{\mp 2i\over
(2\pi\lambda\sqrt{1+e^{2\pi z}})^{1/2}}
e^{\pm i F(\lambda,z)}\cr
\chi^{\pm}_L&\equiv \bigl(\sqrt{k}\mp {i\over \sqrt{k}}\bigr)\psi^+
+\bigl(\sqrt{k}\pm {i\over \sqrt{k}}\bigr)\psi^-
{\buildrel \lambda\to -\infty\over\sim}{\mp 2i\over
(2\pi |\lambda|\sqrt{1+e^{2\pi z}})^{1/2}}
e^{\pm i F(|\lambda|,z)}\cr}
}
where

\eqn\asdfs{
\eqalign{
F(\lambda,z)&={1\over 4}\lambda^2-z log \lambda
+\Phi(z)\cr
\Phi(z)&\equiv {\pi\over 4}+{i\over 4} log \Biggl[
{\Gamma(\half-iz)\over \Gamma(\half+iz)}\Biggr]  \cr
k(z)&=\sqrt{1+e^{2 \pi z}}-e^{\pi z}\cr
k(z)^{-1}&=\sqrt{1+e^{2 \pi z}}+e^{\pi z}\ .\cr}
}

By demanding that the resolvent be properly normalized and vanish at
infinity we see that if we act on $L^2(\IR)$ then we have
\eqn\resolvi{
 \langle\lambda_1 |{1\over H-z}|\lambda_2\rangle=
-{\pi\over 2}\theta(\lambda_1-\lambda_2)
 \chi_R^-(z,\lambda_1) \chi_L^-(z,\lambda_2) +
\lambda_1\leftrightarrow \lambda_2
}
 for $Im(z)>0$.
 From this and the above asymptotics of $\chi$ we obtain \gmfrm.
 If instead we choose a semi-infinite
space $L^2[A,\infty)$ we have for $Im(z)>0$
\eqn\resolvii{
 \langle\lambda_1 |{1\over H-z}|\lambda_2\rangle=-\CO(z)\theta
(\lambda_1-\lambda_2)
 \chi_R^-(z,\lambda_1)\Biggl[\psi^-(z,\lambda_2)-{\psi^-(z,A)\over\psi^+(z,A)}
 \psi^+(z,\lambda_2)\Biggr] +
\lambda_1\leftrightarrow \lambda_2
}
 Normalizing $\CO(z)$ and using the large $\lambda$
asymptotics leads to
the reflection coefficient:
 \eqn\genrefl{
R_q=e^{i\mu log\mu}\mu^{-|q|}i {
(\sqrt{k}- {i\over \sqrt{k}})
-{\psi^-(z,A)\over \psi^+(z,A)}(\sqrt{k}+{i\over \sqrt{k}})
\over
(\sqrt{k}+{i\over \sqrt{k}})
-{\psi^-(z,A)\over \psi^+(z,A)}(\sqrt{k}-{i\over \sqrt{k}})}
\sqrt{\Gamma(\half-i\mu+|q|)\over\Gamma(\half+i\mu-|q|)}
}
where $z = \mu + iq$.

In particular, for $A=0$ we recover the simpler expression \gamhf.

\appendix{B}{An Identity on Gamma Functions}

In \moore\ the four point function was calculated as the limit at small loop
lengths of the macroscopic loop amplitude. The formalism of \moore\ allows
calculations to all orders in the genus expansion.
Comparing our result \twotwo\ to eqn. (4.38) of \moore ,we find that
the first two terms of each (corresponding to the two-bounce contribution) are
equal to within corrections that vanish at any order.
Requiring agreement of the remainder of the two formulas for the amplitude
leads to the rather remarkable result
\eqn\egamident{\eqalign{
&Im\Biggl\{ e^{i\pi S/2}
\sum_{n=0}^\infty {(-1)^n\over n!}{\Gamma(-q_1+n)\over \Gamma(-q_1)}
\biggl({\Gamma(q_3+n)\over \Gamma(q_3)}+{\Gamma(q_4+n)\over \Gamma(q_4)}\biggr)
\cr
&\qquad\qquad \times \biggl(
{\Gamma(\half-i\mu+S-n)\over \Gamma(\half-i\mu)}-
{\Gamma(\half-i\mu+q_1-n)\over\Gamma(\half-i\mu-q_2)}\biggr)
\Biggr\} =\cr\cr
&Im \Biggl\{ e^{i \pi S/2}
\Biggl[ {\Gamma(\half - i\mu + q_2)\over \Gamma(\half - i\mu + q_2 + q_3)}
{\Gamma(\half - i\mu - q_4)\over \Gamma(\half - i\mu) } +
{\Gamma(\half - i\mu + q_2)\over \Gamma(\half - i\mu + q_2 + q_4)}
{\Gamma(\half - i\mu - q_3)\over \Gamma(\half - i\mu) }
\Biggr] \Biggr\} \cr
}}
where $S = \half \sum |q_i|$. This equality holds to all orders in an
asymptotic expansion at large $\mu$. Expanding both sides and equating
the coefficients, which are expressed in terms of polygamma functions, will
lead to identities on these.

\appendix{C}{Small Energy and Topological Expansions}

We write here the small energy expansions of some scattering amplitudes to
all orders in the $1\over \mu$ expansion. These are useful for some explicit
unitarity checks to all orders in the genus expansion, since higher n-point
functions enter the unitarity equations at higher powers of the energy. We
give  here the two-point function to $\CO(q^9)$,  the three-point function
to $\CO(q^6)$, the four-point function to $\CO(q^7)$ in  two kinematic
regimes and the $n$-point
function to $\CO(q^{n+1})$ in the $1\to n$ regime.
We also give the two-point function at genus one, two and three.

The coefficients at each order in
the small energy expansion are written in terms of $\psi_n$ where
\eqn\pgam{\eqalign{
\psi_0 \sim &  log \mu +\sum_{n=1}^\infty {(-1)^nB_{2n}\over
2n}(1-2^{-2n+1})\mu^{-2n}\cr
\psi_n = &\qquad \bigl({d\over d\mu}\bigr)^n \psi_0 \cr }}.

\subsec{Two Point Function}

(a) Small Energy Expansion

\eqn\tptfn{\eqalign{
\mu^q R(q;-q) =
&qe^{q\psi_0} \biggl[ 1 - {q^3\over 12} \psi_2 - {q^4\over 24}
\psi_1^2 + {q^5\over360} \psi_4 + {q^6\over 240} \psi_2^2 \cr
&\qquad +
{q^6\over 180} \psi_1\psi_3 + q^7 ( {7\over 1440} \psi_1^2\psi_2
- {1\over 20160} \psi_6 ) \cr
&\qquad + q^8 ( {1\over1920} \psi_1^4 - {23\over120960} \psi_3^2 -
{19\over60480} \psi_2\psi_4 - {1\over6720} \psi_1\psi_5 ) \cr
&\qquad + q^9
( {-29\over181440} \psi_2^3 -{41\over60480} \psi_1\psi_2\psi_3 -
{1\over5040} \psi_1^2\psi_4 + {1\over1814400} \psi_8 ) + \cdots
\biggr] \cr}}
where $q$ is the euclidean momentum and $q>0$.

\medskip

(b) Genus Expansion

The genus one and two results are written here in a form which emphasizes
that our results can be expanded to give  integrals of $n$ vertex operators
with any choice of charges over moduli spaces of riemann surfaces
of arbitrarily high genus. The answer is always some polynomial in the charges
$q_i$ with coefficients that are a combination of Bernoulli numbers.

 \eqn\mutwo{
\int_{\CM_{1,2}}<\CT_q\CT_{-q}> = -{1\over24}q^2(q-1)(q^2-q-1)}

\eqn\mufo{
\int_{\CM_{2,2}}<\CT_q\CT_{-q}> =
{q^2\over5760}\prod_{r=1}^{3}(q-r)(3q^4-10q^3-5q^2+12q+7)}

\eqn\musix{\eqalign{
\int_{\CM_{3,2}}<\CT_q\CT_{-q}> =& -{q^2\over2903040}
\prod_{r=1}^{5}(q-r)
(9q^6-63q^5+42q^4+217q^3-205q-93)\ .
\cr}}

\subsec{Three Point Function}

\eqn\trptfn{\eqalign{
\mu^{|q_3|} R(q_1,q_2;q_3)=& - e^{|q_3|\psi_0} q_1q_2|q_3|\biggl[\psi_1 -
{1\over12}(q_1^2+q_1q_2+q_2^2)\psi_3-\cr &{1\over 12}|q_3|(2q_2^2 +
 q_1q_2+q_1^2) \psi_1\psi_2 + \cdots\biggr]\ ,\cr}}
where $q_1,q_2>0$ and $q_3<0$.

\subsec{Four Point Function}

The four point function in the $1\to 3$ and
$2\to 2$ kinematic regimes is given up to
$\CO(q^7)$ where we can check that one particle irreducible amplitudes
of \kdf\ are not analytic at genus one.

\eqn\fptfna{\eqalign{
\mu^{|q_4|} R(q_1,q_2,q_3;q_4) = &e^{|q_4|\psi_0}q_1q_2q_3|q_4|\biggl[\psi_2
+|q_4|\psi_1^2-{1\over24}(q_1^2+q_2^2+q_3^2+q_4^2)\psi_4 -\cr &\qquad
{\psi_2^2\over
2}({q_1^3\over3}+{q_1^2q_2\over2}+{q_1q_2^2\over2}+{q_2^3\over3}+
{q_1^2q_3\over2
}+{q_1q_2q_3\over2}+\cr  &\qquad {q_2^2q_3\over2}+{q_1q_3^2\over2}+
{q_2q_3^2\over2}+
{q_3^3\over3})-\cr  &\qquad {\psi_1\psi_3\over2}
({q_1^3\over2}+{5q_1^2q_2\over6}+
{5q_1q_2^2\over6}+{q_2^3\over2}+{5q_1^2q_2\over6}
+{5q_1^2q_3\over6}+ \cr &\qquad
 {q_1q_2q_3}
+{5q_2^2q_3\over6}+{5q_1q_3^2\over6}+{5q_2q_3^2\over6}+{q_3^3\over2})+\cdots
\biggr]\cr. }}

In the following $q_1=max{|q_i|}$

\eqn\fptfnb{\eqalign{
\mu^{q_1+q_2} R(q_1,q_2;q_3,q_4)= &e^{|q_4|\psi_0}q_1q_2|q_3||q_4|
\biggl[\psi_2-
q_1\psi_1^2-\cr &\qquad
{\psi_4\over12}(q_1^2+q_1q_2+q_2^2+q_1q_3+q_2q_3+q_3^2)
\cr &\qquad
-{\psi_2^2}({q_1^3\over6}+{q_1^2q_2\over4}+{q_1q_2^2\over4}+{q_2^3\over12}+
{q_1^2q_3\over4}+{q_1q_2q_3\over4}+{q_1q_3^2\over4})
\cr &\qquad - \psi_1\psi_3( {q_1^3\over4}+{q_1^2q_2\over4}+{q_1q_2^2\over4}+
{q_2^3\over12}+{q_1^2q_3\over4} ) + \cdots \biggr] \ .\cr}}

\subsec{N-point Function}

For $n>2$ we have:
\eqn\nptfn{\eqalign{
\mu ^{|q_n|} R(q_1,q_2,\dots;q_n) =&  q_1q_2 \dots q_{n-1} |q_n|\biggl[
 \psi_{n-2}\cr
+& {|q_n|\over 2}({\sum_{r=0}^{n-2}} {n-2\choose r}
\psi_{n-2-r}\psi_r )
+ \cdots \biggr]\ .\cr} }

These first two terms can easily be obtained by using the recursion relation

\eqn\rec{
R(q_1,q_2,\dots;q_n)\sim q_1 {\partial\over\partial \mu}R(q_2,\dots; q_n)
}
for $q_1\to 0,$ and they are precisely what we expect from unitarity.

\appendix {D}{Proof of Equivalence to Bosonization}

We wish to prove that the $S$-matrix we compute may be obtained from the
simple scattering amplitude \sff\ for free fermions via the bosonization
prescription of \bosnztion. We will construct the proof in Euclidean space to
connect to the derivation of the `filtration formula'. As with that formula,
all results are later continued to Minkowski space.
Formally, we need to show
\eqn\bozinc{\eqalign{
&\int_{-\infty}^\infty \prod_id\xi_i
\prod_{i=1}^n a(\mu+\xi_i)a^\dagger(\mu+\xi_i+q_i)|\mu\rangle = \cr
&\qquad \sum_{k = 1}^n \sum_{T_1 \amalg \ldots \amalg T_k = S^-}
\prod_{j = 1}^k \int\, d\xi_j f_-(T_j,-\xi_j)
\prod_{j = 1}^k a^{\dagger} \big( \mu + \xi_j + q(T_j) \big)
a\big( \mu + \xi_j \big) \, |\mu \rangle \ , \cr
}}
where $S^- = \{q_1,\ldots q_n \}$.
Note that since
$f_-(T,Q) \propto - \theta (-Q) \theta \big(Q-q(T)\big)$, the
left-hand side of \bozinc\ is normal-ordered.
The proof proceeds by induction on $n$. For $n = 1$, the claim is
explicitly verified in \bosnztion.
Assuming the assertion holds for $n-1$ tachyons, we proceed to $n$ by
multiplying the ``fermionized'' state corresponding to
$S=\{q_2,\ldots q_n\}$
on the left by
$\int\, d\xi_1 a (\mu + \xi_1) a^{\dagger}(\mu + \xi_1 + q_1)$.
To do this, we split the integral into a
normal-ordered part $I_1 = \int_0^{-q_1} d\xi_1 $ and the remainder $I_2$.
In $I_1$, the operators can be commuted to the right to canonically order the
result, leading to
\eqn\iione{
I_1 = \sum_{k = 1}^{n-1}
\sum_{{\scriptstyle T_1 \amalg \ldots \amalg T_k = S^-}
\atop {\scriptstyle T_k = \{q_1\} }}
\int \prod_j d\xi_j
\prod_{j = 1}^k f_-(T_j,-\xi_j) a^{\dagger} \big( \mu + \xi_j + q(T_j) \big)
a\big( \mu + \xi_j \big) \, |\mu \rangle \ .
}
In $I_2$, the operators on the left annihilate $|\mu\rangle$, so the
contribution is a sum of anticommutator terms. Using the identity
\eqn\identfmin{
f_-(T,Q) = f_-(T\setminus \{q^*\}, Q - q^* ) - f_-(T\setminus \{q^*\},Q )
}
for $q^* \in T$, one verifies that the $j$\/th contribution is
\eqn\combrak{\eqalign{
&\left( \int_{-\infty}^0 \! + \int_{-q_1}^{\infty} \right)d\xi_1
\Big[ a(\mu + \xi_1) a^{\dagger} (\mu + \xi_1 + q_1),
a^{\dagger} \big(\mu + \xi_j + q(T_j) \big)
a \big( \mu + \xi_j \big) \Bigl] = \cr
&\qquad
f_-(T_j\amalg \{q_1\},-\xi_j)
a^{\dagger} \big(\mu + \xi_j + q(T_j) + q_1 \big)
a \big( \mu + \xi_j \big) \ .\cr
}}
Adding the two contributions proves the claim by induction.

\listrefs
\listfigs

\bye